\documentclass[10pt,english,aps,prd,urlcolor=blue,linkcolor=black,nofootinbib, notitlepage,onecolumn, groupedaddress]{revtex4-1}

\usepackage[utf8]{inputenc}
\usepackage{babel}
\usepackage{graphicx}
\usepackage{amsmath,amsfonts,amssymb,amsthm}
\usepackage{verbatim}
\usepackage{bm}
\usepackage{bbm}
\usepackage[]{natbib}
\usepackage{url}
\usepackage{slashed}
\usepackage{upgreek}
\usepackage{enumitem}
\usepackage{epsf,epsfig}
\usepackage{graphics}
\usepackage{tabularx,makecell}
\setcellgapes{2.5pt}

\usepackage[colorlinks,citecolor=black]{hyperref}

\usepackage{mathtools}	
\usepackage{braket} 	
\usepackage{subcaption} 	
\usepackage{cancel} 	
\usepackage{mathrsfs}	
\usepackage{setspace}	
\usepackage{dsfont} 	
\usepackage{tensor} 	
\usepackage{esint}		
\usepackage[all]{xy}	
\usepackage{stmaryrd}	
\usepackage{extarrows}	
\usepackage{siunitx} 	
\usepackage{feynmp-auto}		
\usepackage{tikz}		
\usepackage{enumerate} 	
\usepackage{eepic}      
\usepackage{hyperref}
\usepackage{bookmark}

\renewcommand{\d}{\mathrm{d}}

\newtheoremstyle{mythm}
{}
{}
{\slshape}
{}
{\bfseries\sffamily}
{.}
{ }
{}
\newtheoremstyle{mydef}
{}
{}
{}
{}
{\bfseries\sffamily}
{.}
{ }
{}

\theoremstyle{mythm}

\theoremstyle{mydef}

\numberwithin{equation}{section}

\newcommand{\ad}{\operatorname{ad}}

\DeclareMathOperator{\tr}{tr}

\def\i{\mathrm{i}}
\def\e{\mathrm{e}}

\newcommand{\D}{\mathrm{d}}
\newcommand{\rom}[1]{\uppercase\expandafter{\romannumeral #1\relax}}

\DeclareMathOperator*{\SumInt}{%
\mathchoice%
  {\ooalign{$\displaystyle\sum$\cr\hidewidth$\displaystyle\int$\hidewidth\cr}}
  {\ooalign{\raisebox{.14\height}{\scalebox{.7}{$\textstyle\sum$}}\cr\hidewidth$\textstyle\int$\hidewidth\cr}}
  {\ooalign{\raisebox{.2\height}{\scalebox{.6}{$\scriptstyle\sum$}}\cr$\scriptstyle\int$\cr}}
  {\ooalign{\raisebox{.2\height}{\scalebox{.6}{$\scriptstyle\sum$}}\cr$\scriptstyle\int$\cr}}
}

\begin{document}
\title{Gradient effects on false vacuum decay in gauge theory}
\author{Wen-Yuan Ai}
\email[E-mail:  ]{wenyuan.ai@uclouvain.be}
\affiliation{Centre for Cosmology, Particle Physics and Phenomenology,\\ Université catholique de Louvain, Louvain-la-Neuve B-1348, Belgium}
\author{Juan S. Cruz}
\email[E-mail:  ]{juan.cruz@tum.de}
\author{Bj\"orn~Garbrecht}
\email[E-mail:  ]{garbrecht@tum.de}
\author{Carlos Tamarit}
\email[E-mail:  ]{carlos.tamarit@tum.de}
\affiliation{Physik Department T70, James-Franck-Stra{\ss}e,\\
	Technische Universit\"at M\"unchen, 85748 Garching, Germany}
\date{\today}

\thispagestyle{empty}

\begin{minipage}{.5\textwidth}
 \begin{flushleft}
{
\small
CP3-20-21
}
\end{flushleft}
\end{minipage}\begin{minipage}{.5\textwidth}
\begin{flushright}
{
\small
TUM-HEP-1265-20
}
\end{flushright}
\end{minipage}

\begin{abstract}
{We study false vacuum decay for a gauged complex scalar field in a polynomial potential with nearly degenerate minima. Radiative corrections to the profile of the nucleated bubble as well as the full decay rate are computed in the planar thin-wall approximation using the effective action. This allows to  account for the inhomogeneity of the bounce background and the radiative corrections in a self-consistent manner. In contrast to scalar or fermion loops, for gauge fields one must deal with a coupled system that mixes the Goldstone boson and the gauge fields, which considerably complicates the numerical calculation of Green's functions. In addition to the renormalization of couplings, we employ a covariant gradient expansion in order to systematically construct the counterterm for the wave-function renormalization. The result for the full decay rate however does not rely on such an expansion and accounts for all gradient corrections at the chosen truncation of the loop expansion. The ensuing gradient effects are shown to be of the same order of magnitude as non-derivative one-loop corrections.}  
\end{abstract}
\maketitle

\tableofcontents

\section{Introduction}

With the discovery of the Higgs boson in 2012, the last missing piece of the Standard Model (SM) was set into place~\cite{Aad:2012tfa,Chatrchyan:2012xdj}. Remarkably, its properties appear to lie precisely in the narrow parameter range where the SM could, in principle, be a consistent effective field theory for energies up to the Planck scale. If the Higgs quartic coupling were slightly larger, the SM would exhibit a Landau pole at energies below the Planck scale that would severely limit the prediction power of the theory. If it were slightly smaller, the electroweak vacuum would be too short-lived. This criticality may have very important implications for possible extensions of the SM. 

Taken at face value, the best-fit parameters of the SM indicate that the electroweak vacuum is metastable. This is due to the running of the Higgs quartic, which turns negative at an energy scale much larger than the electroweak one, at around $10^{11}\,{\rm GeV}$, inducing a lower-lying, true vacuum in the effective potential at high field values. The electroweak vacuum may then decay to this global minimum through quantum tunnelling. The scalar potential, the top Yukawa and the electroweak gauge couplings have been extracted from data at full two-loop next-to-next-leading order (NNLO) precision~\cite{Degrassi:2012ry, Buttazzo:2013uya}. These parameters have been extrapolated to large energies using the full three-loop renormalization-group-equation (RGE) to NNLO precision~\cite{Buttazzo:2013uya}. With these calculations, the Higgs and top-quark masses of $125\,{\rm GeV}$ and $173\,{\rm  GeV}$ respectively, suggest a lifetime for the electroweak vacuum that is longer than the age of our Universe, leading to the metastability scenario~\cite{EliasMiro:2011aa,Degrassi:2012ry}. However, in comparison to the running couplings, the radiative corrections that appear in the corresponding tunnelling problem have been computed less accurately. So far the one-loop radiative corrections to the decay rate due to fluctuations about the classical bounce have been calculated for the SM in references~\cite{Isidori:2001bm, Branchina:2014rva, DiLuzio:2015iua, Chigusa:2017dux, Andreassen:2017rzq}. 

The tunneling rate is sensitive to the solitonic field configuration known as the ``bounce''~\cite{Coleman:1977py, Callan:1977pt}. The equation of motion for the latter is often derived from a renormalization-group-improved scalar \emph{potential}, with the running coupling constants 
evaluated at a scale given by the typical values of the scalar field in the bounce 
solution. Yet the bounce is an inhomogeneous configuration whose equation of motion should be determined by the \emph{effective action}. 

The beta functions for the couplings, or more generally the Coleman--Weinberg effective potential~\cite{Coleman:1973jx}, do not account for the  gradient effects arising from the inhomogeneity of the background. To date, the latter have only been accounted for in the calculation of the lifetime of the SM electroweak vacuum when estimating the one-loop fluctuation determinants around the bounce, but not when obtaining the bounce itself~\cite{Isidori:2001bm, Branchina:2014rva, DiLuzio:2015iua, Chigusa:2017dux, Andreassen:2017rzq}. The  one-loop effective action associated with the fluctuation determinants is often computed using the Gel'fand--Yaglom method. While this is a powerful approach to obtain this quantity either analytically or numerically, it is not clear how to extend it beyond one-loop order. 

In order to advance systematically in accuracy, one may pursue an expansion of the effective action and the equations of motion consistent with it in terms of \emph{Green's functions}. This Green's function approach has been carried out in models with interactions among scalar fields~\cite{Garbrecht:2015oea,Garbrecht:2015yza} as well as for Yukawa theory~\cite{Ai:2018guc}. These studies show that the gradient corrections to the one-loop result are comparable with terms that appear at two-loop order in the case of a quasi-degenerate quartic potential. Nonetheless, the impact of the self-consistent Green's function approach is expected to be larger when the scalar potential is nearly scale-invariant, as it is the case in the SM. In this situation the spontaneous breaking of the approximate dilatational symmetry of the bounce gives rise to a pseudo-Goldstone mode. In the approximation of a scale-invariant bounce, the path integration over the Goldstone mode can be traded for a collective coordinate, where different methods of evaluating the integral over the latter have been proposed~\cite{Chigusa:2017dux, Andreassen:2017rzq}. On the other hand, when appreciating that the self-consistently obtained bounce itself breaks scale invariance, the functional determinant can be evaluated without transforming the path integral and regulating it ad-hoc. The self-consistent computation of the effective action and the resulting bounce can then be understood as a summation of one-loop diagrams. It leads to infrared effects giving logarithms which dominate one-loop corrections to the effective action in a model consisting of only scalar fields~\cite{Garbrecht:2018rqx}. The self-consistent computation of these one-loop contributions and the quantum-corrected bounce therefore remains an important task to be addressed in the SM.

At present, such a full calculation of the tunnelling rate in the SM to next-to-leading order (NLO) accuracy (i.e. including the infrared logarithms that appear at that order by a self-consistent computation of the bounce) requires further methodical development. Considering the important role played by the W and Z~bosons in electroweak vacuum metastability, we extend here the methods developed in Refs.~\cite{Garbrecht:2015oea, Ai:2018guc} to 
gauge theories. Although our work elaborates on a specific model and is not applied to realistic phenomenology, we aim to provide an example for how to include gradient effects on the decay rate of the false vacuum through a self-consistent scheme in gauge theories. Further technical developments and
the application to tunnelling in the SM are left for future work.

This paper is organized as follows. In Section~\ref{sec:CCformalism} we review the Callan--Coleman formalism for false vacuum decay as well as the general way of calculating radiative corrections to false vacuum decay at higher-order using the effective action. This is followed in Section~\ref{sec:Setup} by the application of the effective action method to false vacuum decay in gauge theory. In contrast to pure scalar and Higgs--Yukawa models, for gauge theories there is a coupled sector involving the fluctuations of the gauge and Goldstone bosons which demands a more intricate treatment.  The details of these computations for the $R_\xi$-type gauges are presented in the subsequent two sections. Section~\ref{sec:solvingA4G} is devoted to a particular choice of gauge in which the mixing between the gauge and Goldstone degrees of freedom simplifies. Then, Section~\ref{sec:ren} contains the details of the renormalization procedure applied for this model. The computer implementation and numerical results are reported in Section~\ref{sec:num}. Comments and discussion of the results are given in Section~\ref{sec:Conc}. Throughout this paper, we use $c=1$ and repeated Greek indices at the same level are summed up with Euclidean signature.

\section{Callan--Coleman formalism and the bounce}
\label{sec:CCformalism}

\begin{figure}
  \includegraphics[scale=0.6]{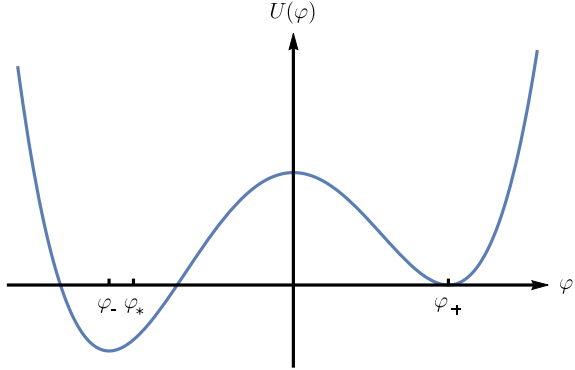}
\captionsetup{justification=raggedright,singlelinecheck=false}
  \caption{The classical  potential $U(\Phi)$ for the archetypical scalar model with false vacuum decay, given by Eqs.~\eqref{eq:themodel} 
  and~\eqref{eq:classicalpotential}.}
  \label{fig:potential}
\end{figure}

In this section, we review the pertinent details of the calculation of the decay rate of a metastable vacuum state following Callan and Coleman~\cite{Coleman:1977py,
Callan:1977pt} and considering the following archetypal model, 
\begin{align}
\label{eq:themodel}
\mathcal{L}_{\rm M}=\frac{1}{2}\eta^{\mu\nu}(\partial_\mu\Phi)(\partial_\nu\Phi) - U(\Phi),
\end{align}
where $\mu,\nu=0,...,3$, $\eta^{\mu\nu}$ is the Minkowski metric with signature $+,-,-,-$ and
\begin{align}
\label{eq:classicalpotential}
U(\Phi) = -\frac{1}{2}\,\mu^2\,\Phi^2 + \frac{1}{3!}\,\lambda_3\,\Phi^3 + \frac{1}{4!}\,\lambda\,\Phi^4 + U_0.
\end{align}
The couplings $\mu^2, \lambda, \lambda_3$ all take positive values, and the cubic term breaks the $\mathbb{Z}_2$ symmetry at tree level such as to lift the degeneracy between the vacua. The potential is assumed to have two minima at $\varphi_+$ and $\varphi_-$, corresponding to the false and the true vacuum, respectively. For convenience, one can choose the constant $U_0$ such that the false vacuum has vanishing energy density. An example  potential is depicted in Figure~\ref{fig:potential}.

In order to obtain the decay rate, Callan and Coleman consider the following {\it Euclidean} false vacuum to false vacuum transition amplitude 
\begin{align}
\label{eq:partition}
Z[0] = \langle \varphi_+| \e^{-H\mathcal{T}/\hbar}|\varphi_+\rangle = \int\mathcal{D}\Phi\:\e^{-\frac{1}{\hbar}S_{\rm E}[\Phi]},
\end{align}
where $H$ is the full Hamiltonian and $\mathcal{T}$ is the amount of Euclidean time 
taken by the transition. The classical Euclidean action is $S_{\rm E}[\Phi]$, which can be 
obtained by a Wick rotation, i.e. $S_{\rm E}=-\i S_{\rm M}(x_0\rightarrow -\i x_4)$ with $x_4\equiv 
\tau$ being the Euclidean time. Written explicitly,
\begin{align}
S_{\rm E}[\Phi]=\int \d^4x \left[\frac{1}{2}\delta^{\mu\nu}(\partial_\mu\Phi)\partial_\nu\Phi+U(\Phi)\right],
\end{align}
where $\delta^{\mu\nu}$ is the Kronecker symbol and $\mu,\nu=1,...,4$ for Euclidean space. Observe that in the Euclidean action the sign in front of the 
potential is flipped. An important consequence is that, for the boundary 
conditions of interest, the Euclidean action allows for a classical soliton 
solution, the so-called \emph{bounce}, as we describe below.  

One can extract the ground-state energy $E_0$ of the system by inserting a complete set of energy eigenstates into the partition functional, i.e.
\begin{align}
\label{eq:partition2}
\langle\varphi_+|\e^{-H\mathcal{T}/\hbar}|\varphi_+\rangle = \sum\limits_n\,\e^{-E_n\mathcal{T}/\hbar}\:\langle\varphi_+|n\rangle\langle n|\varphi_+\rangle,
\end{align}
where $E_0$ has the smallest real part among all $E_n$,
and then taking the limit $\mathcal{T}\rightarrow\infty$ such that the contribution corresponding to $E_0$ dominates. In the case of an unstable state, which can be modelled as a non-normalizable eigenstate with complex energy $E_{0+}$, and with ${\rm Im}E_{0+}$ related to the tunnelling rate, one can isolate the contribution from $E_{0+}$ by appropriately constraining the path integration \cite{Andreassen:2016cvx,Ai:2019fri}. The constraint is enforced by performing the path integration using the method of steepest descent around bounce and multi-bounce saddle-points that can be approximated as combination of the single bounce $\varphi$ in the dilute-gas approximation.

The tree-level bounce is a solution to the classical Euclidean equation of motion
\begin{align}
- \partial^2\varphi + U'(\varphi) = 0
\label{eq:eomBounce}
\end{align}
that satisfies the boundary conditions $\varphi|_{x_4\rightarrow\pm\infty} = 
\varphi_+$ and $\dot{\varphi}|_{x_4=0}=0$, where the dot denotes the derivative with 
respect to $x_4$. The prime denotes the derivative of the classical potential from 
Eq.~\eqref{eq:classicalpotential} with respect to the field $\varphi$. Notice that we are interested in a field configuration that starts \emph{and ends} in the false vacuum, hence its name. For the bounce action to be finite, we also require that 
$\varphi|_{|\bf{x}|\rightarrow\infty} = \varphi_+$. 
Given the anticipated $O(4)$ invariance of the bounce, it is convenient to work in 
four-dimensional hyperspherical coordinates, in which the equation of motion takes 
the form
\begin{align}
-\frac{\d^2\varphi}{\d r^2} - \frac{3}{r} \frac{\d\varphi}{\d r} + U'(\varphi) = 0,
\label{eq:eomExpanded}
\end{align}
with $r^2={\bf x}^2+x_4^2$. The boundary conditions become $\varphi|_{r\rightarrow 
\infty}=\varphi_+$. The solution must be regular at the origin, and we therefore 
require that $\d\varphi/\d r|_{r\,=\,0}=0$. Its form is that of a soliton that 
interpolates between the field value $\varphi_*$ corresponding to the escape point (which lies close to the true vacuum $\varphi_-$) at the origin of Euclidean space and the false vacuum 
$\varphi_+$ at infinity, see Figure~\ref{fig:potential}. Therefore, it describes a four-dimensional hyperspherical bubble nucleated within the false vacuum. This {\it 
classical} solution will be denoted as $\varphi_b$. In the limit in which the potential energy of the true and false vacua become degenerate, it can be argued that the bubble is very thin compared to its large radius~\cite{Coleman:1977py}. This corresponds to the ``thin-wall'' limit. One then may also approximate the bubble wall by a planar configuration.

When evaluated at the bounce, the fluctuation operator possesses a negative eigenvalue,
and naively performing the Gaussian integral produces a divergent result. A physically meaningful answer, however, can be found through careful analytic continuation by which one obtains an imaginary part of the energy, which is interpreted in terms of the complex energy $E_{0+}$ of the false vacuum state. In terms of $Z[0]$, the 
decay rate is given by~\cite{Callan:1977pt}
 \begin{align}
\gamma=\frac{2\:|\textrm{Im}\,Z[0]|}{\mathcal{T}}.
\label{eq:decayrate2}
\end{align} 
Note that in the above formula, the partition function is to be evaluated by expanding around the bounce solution and normalized to be one when evaluated at the false vacuum. At one-loop order, one has~\cite{Callan:1977pt}
\begin{align}
\frac{\gamma}{V}=\left(\frac{B}{2\pi\hbar}\right)^2\left|\frac{\det^\prime[-\partial^2+U''(\varphi_b)]}{\det[-\partial^2+U''(\varphi_+)]}\right|^{-1/2}\,\e^{-B/\hbar},
\label{eq:decay-oneloop}
\end{align}
where $B=S_{\rm E}[\varphi_b]$ is the bounce action, $\det'$ means that the zero 
eigenvalues are to be omitted from the determinant and a prefactor $\sqrt{B/2\pi\hbar}$ is included for each of the four collective coordinates that correspond to spacetime translations~\cite{GervaisSakita1975}.

In order to reformulate Eq.~\eqref{eq:decayrate2} in a way that radiative effects can be systematically considered, we make use of the effective 
action~\cite{Jackiw:1974cv,Cornwall:1974vz}. For conciseness, we employ the DeWitt notation 
\begin{align}
J_x\varphi_x = \int \d^4x\; J(x)\,\varphi(x),
\end{align}
in which repeated continuous indices are integrated over. Recall that the effective 
action is defined as the Legendre transform
\begin{align}
\Gamma[\varphi]& = -\hbar\log Z[J] + J_x\varphi_x,
\label{eq:effectiveaction1}
\end{align}
where 
\begin{align}
Z[J] & = \int\!\mathcal{D}\Phi \;\exp\,\left[-\frac{1}{\hbar}(S[\Phi] - J_x\Phi_x)\right]
\label{eq:partitionfunctional}
\end{align} 
and 
\begin{align}
\varphi_x& = \langle\Omega|\Phi_x|\Omega\rangle|_{J} = \:\hbar\frac{\delta\log 
Z[J]}{\delta J_x}
\label{eq:onePointphi}
\end{align}
is the one-point function in the presence of the source $J$. From the effective action, one obtains the equation of motion
\begin{align}
\frac{\delta\Gamma[\varphi]}{\delta\varphi_x} = J_x,
\label{eq:point1}
\end{align}
which gives the quantum-corrected bounce. The effective action thus provides a quantum version of the principle of least action.

In terms of the effective action, the decay rate~\eqref{eq:decayrate2} can be 
written as~\cite{Garbrecht:2015oea, Garbrecht:2015cla, Garbrecht:2015yza, 
Plascencia:2015pga}
\begin{align}
\gamma= \frac{2\, |{\rm Im}\, \e^{-\Gamma[\varphi]/\hbar}|}{\mathcal{T}},
\label{eq:decay3}
\end{align}
where the quantum-corrected bounce $\varphi$ is the solution to the quantum equation of motion~\eqref{eq:point1} with $J=0$. In case the vacuum structure is generated by radiative corrections in the first place, one can deal with it using the two-particle irreducible effective action and evaluate
the partition function by expanding about the self-consistent solution to the quantum-corrected one and two-point functions, as explained in Refs.~\cite{Garbrecht:2015cla,Garbrecht:2015yza,Millington:2019nkw}.

\section{False vacuum decay in gauge theory}
\label{sec:Setup}

Our goal is to carry out a proof-of-principle calculation of the decay rate of the false vacuum in an Abelian gauge theory, including the effects of radiative corrections on the self-consistent bounce solution and accounting for gradient effects  without resorting to a gradient expansion of the effective action. For this purpose, we study a model with the following particle content: a complex scalar field $\Phi$, a $U(1)$ gauge field $A_\mu$ and the associated ghost fields, $\eta$ and $\bar{\eta}$, with the following Euclidean Lagrangian 
\begin{align}
\label{eq:Lagrangian}
\mathcal{L}=(\partial_{\mu}\Phi^\star+\i gA_{\mu}\Phi^\star)
(\partial_{\mu}\Phi-\i gA_{\mu}\Phi)+U(\Phi^* \Phi)+\frac{1}{4}\,F_{\mu\nu}F_{\mu\nu}+\mathcal{L}_{\rm G.F.}+\mathcal{L}_{\rm ghost},
\end{align}
where $F_{\mu\nu}=\partial_\mu A_{\nu}-\partial_\nu A_{\mu}$. Here $U(\Phi^* \Phi)$ is the scalar potential while $\mathcal{L}_{\rm G.F.}$ and $\mathcal{L}_{\rm ghost}$ are the 
gauge fixing and Faddeev-Popov ghost terms, respectively. Since the cubic term 
in Eq.~\eqref{eq:classicalpotential} for the archetypical real scalar model~\eqref{eq:themodel} is not allowed by the gauge symmetry, one may attempt to use a potential
\begin{align}
\label{eq:higgspotential}
U(\Phi)\ =\ -\:\,\mu^2\,|\Phi|^2\:+\:{\lambda}\,|\Phi|^4
\end{align}
resembling the case of the SM Higgs field. But this theory simply displays spontaneous symmetry breaking (SSB) via the Higgs mechanism while not having 
a metastable vacuum. 

\begin{figure}[htbp]
	\centering
	\includegraphics[clip,scale=0.5]{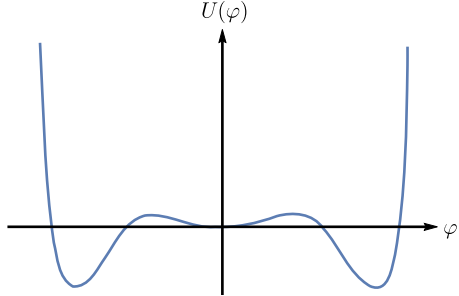}
\captionsetup{justification=raggedright,singlelinecheck=false}
	\caption{Tree level potential $U(\Phi^* \Phi)$ having different vacua (local or global minima).}
	\label{fig:potentialDeg}
\end{figure}

Instead, we here consider the following potential:
\begin{align}
\label{eq:model}
U(\Phi)\ =\ \:\,\alpha\,|\Phi|^2\:+\:{\lambda}\,|\Phi|^4\;+\:{\lambda_6}\,|\Phi|^6.
\end{align}
The false vacuum is still located at $\Phi=0$. The last term of dimension six allows us to manufacture two vacua for a certain region of parameters, such that the potential presents a profile as in Figure~\ref{fig:potentialDeg}.
Phenomenologically, this term can be understood as an effective operator induced by physics beyond the SM (BSM), which is suppressed by the energy scale of new physics. The model specified through Eqs.~(\ref{eq:Lagrangian}) and~(\ref{eq:model}) can arise, e.g., from an ultraviolet (UV) completion with extra heavy fermions coupling to $\Phi$ through Yukawa interactions, and whose associated loop corrections generate higher-dimensional contributions in the scalar potential, these arguments are made more precise in Sec.~\ref{sec:num} for our particular set of parameters. Alternatively, one can also consider loop corrections from a singlet scalar $S$ with an $S^2|\Phi|^2$ interaction. If a coupling $S|\Phi|^2$ is allowed, then one can also generate the higher-dimensional terms from tree-level diagrams with heavy scalar propagators.  All these approaches generate a $\Phi^6$ term plus higher-order interactions, which are suppressed by increasing powers of the  coupling between the heavy fields and $\Phi$, times the inverse of the heavy mass. Once we have specified a parametric benchmark point, we will comment on its validity from an effective field theory point of view. The difference between the model~\eqref{eq:model} and the $\lambda\Phi^4$ model is that the model~\eqref{eq:model} allows for false and true vacua at tree-level.

When considering the instability of the Higgs field in the SM, the quadratic term is often neglected and the renormalization scale is chosen to be the Higgs instability scale, leading to a negative quartic coupling.  The potential then reduces to a quartic monomial with negative $\lambda$, such that the false vacuum is $\Phi=0$, and the false and true vacua are far from degenerate. (In fact, there is no true vacuum unless radiative corrections eventually turn $\lambda$ positive again for very large field values, as is the case in the SM.) While the situation in the latter is of ultimate interest,  here we develop the method of Green's functions in the gauge sector by considering a simpler model that leaves aside for now the issues coming from the approximate scale-invariance of the SM. The question of how to deal with scale-invariance in the Green's function approach is partly addressed for a model consisting only of scalar fields in Ref.~\cite{Garbrecht:2018rqx}.

In the model given by the potential in Eq.~\eqref{eq:model} it is not possible to obtain a general analytic expression for the bounce solution  $\varphi_b$ to the classical Euclidean equation of motion. Moreover, it will prove convenient to use as an initial approximation to the quantum bounce $\varphi$ a configuration $\varphi_0$ that solves the equations of motion that follow from using the one-loop Coleman-Weinberg potential rather than its tree-level counterpart. We obtain $\varphi_0$ as a numerical solution. Additionally, we use the thin-wall approximation---as mentioned before, valid when the minima are quasi-degenerate---and the planar-wall limit, in which the bubble becomes infinitely large and its $O(4)$ symmetry can be traded for  $O(3)$, such that the bounce solution becomes invariant under translations parallel to the bubble wall, which can be taken to be orthogonal to the $x_4$-axis.

\subsection{Effective action}

To obtain the decay rate along the lines of Section~\ref{sec:CCformalism},
we first work out the effective action~\cite{Jackiw:1974cv,Cornwall:1974vz} for the gauge theory~\eqref{eq:Lagrangian}. The Euclidean partition function for this case is 
\begin{align}
Z[J,K_\mu,\bar{\psi},\psi]=\int\mathcal{D}\Phi\mathcal{D}
A_{\mu}\mathcal{D}\eta\mathcal{D}\bar{\eta}\,
\e^{-\frac{1}{\hbar}\int \D^4x [\mathcal{L}-J(x)\Phi(x)-K_\mu(x) A_\mu(x)-\bar{\psi}(x)\eta(x)-\bar{\eta}(x)\psi(x)]},
\end{align}
where we have introduced ghost fields $\eta,\bar\eta$, while $J,K_\mu,\bar{\psi},\psi$ are the external currents corresponding to the various fields.
Defining the one-point expectation values in the presence of sources as
\begin{align}\label{eq:onepointfunctions}
\begin{aligned}
&\varphi_x=\langle\Omega|\Phi_x|\Omega\rangle|_{J,K_\mu,\bar{\psi},\psi} =\hbar\:\frac{\delta\log Z[J,K_\mu,\bar{\psi},\psi]}{\delta J_x},\\
&\mathcal{A}_{\mu;x}=\langle\Omega|A_{\mu;x}|\Omega\rangle|_{J,K_\mu,\bar{\psi},\psi}=\hbar\:\frac{\delta\log Z[J,K_\mu,\bar{\psi},\psi]}{\delta K_{\mu;x}},\\
&\bar{H}_x=\langle\Omega|\bar{\eta}_x|\Omega\rangle|_{J,K_\mu,\bar{\psi},\psi}=-\ \hbar\:\frac{\delta\log Z[J,K_\mu,\bar{\psi},\psi]}{\delta \psi_x},\\
&H_x=\langle\Omega|\eta_x|\Omega\rangle|_{J,K_\mu,\bar{\psi},\psi}=\hbar\:\frac{\delta\log Z[J,K_\mu,\bar{\psi},\psi]}{\delta \bar{\psi}_x},
\end{aligned}
\end{align}
the effective action is the Legendre transform of the partition function,
\begin{align}
\label{eq:effectiveAction0}
\Gamma[\varphi,\mathcal{A}_{\mu},\bar{H},H]=-\hbar\,\log Z[J,K_\mu,\bar{\psi},\psi]+J_x\,\varphi_x+K_{\mu;x}\,\mathcal{A}_{\mu;x}+\bar{H}_x\,\psi_x+\bar{\psi}_x\,H_x.
\end{align}
It then follows that
\begin{align}
\label{eq:currents}
\begin{aligned}
\frac{\delta\Gamma[\varphi,\mathcal{A}_{\mu},\bar{H},H]}{\delta\varphi_x}&=J_x, &
\frac{\delta\Gamma[\varphi,\mathcal{A}_{\mu},\bar{H},H]}{\delta\mathcal{A}_{\mu;x}}&=K_{\mu;x},\\
\frac{\delta\Gamma[\varphi,\mathcal{A}_{\mu},\bar{H},H]}{\delta\bar{H}_x}&=\psi_x, &
\frac{\delta\Gamma[\varphi,\mathcal{A}_{\mu},\bar{H},H]}{\delta H_x}&=-\bar{\psi}_x.
\end{aligned}
\end{align}

For false vacuum decay in the present model, the bounce corresponds to the scalar one-point expectation value, whilst the one-point expectation values for other fields remain zero. We therefore abbreviate $\Gamma[\varphi,\mathcal{A}_\mu=\bar{H}=H=0]$ as $\Gamma[\varphi]$ and $S[\varphi,\mathcal{A}_\mu=\bar{H}=H=0]$ as $S[\varphi]$.
The tunnelling rate~\cite{Garbrecht:2015oea, Garbrecht:2015cla, Garbrecht:2015yza, Plascencia:2015pga, Ai:2018guc} is then
\begin{align}
\label{decayrate-n-loop}
\frac{\gamma}{V}=\frac{2\,|{\rm Im}\,\e^{-\Gamma^{(n)}[\varphi^{(n)}]/\hbar}|}{V{\cal T}},
\end{align}
where $\Gamma^{(n)}[\varphi^{(n)}]$ and $\varphi^{(n)}$ are the effective action and corrected bounce at $n$-loop order
\begin{align}
\label{eq:quantumeom}
\left.\frac{\delta\Gamma^{(n)}[\varphi]}{\delta\varphi_x}\right|_{\varphi_x=\varphi_x^{(n)}}=0
\end{align}
with $O(4)$-symmetric boundary conditions $\varphi^{(n)}|_{|x|\rightarrow \infty}=0$. These are the equations \eqref{eq:decay3} and \eqref{eq:point1} from the previous section, applied to the present case study. In what follows we calculate the corrected bounce $\varphi^{(1)}$ at one-loop order and the resulting quantum-corrected decay rate when one substitutes $\varphi^{(1)}$ into $\Gamma^{(1)}$ in Eq.~\eqref{decayrate-n-loop}.

We now expand the quantum field $\Phi$ around the classical bounce background 
$\varphi_b=\varphi^{(0)}$ in terms of real fields,
\begin{align}
\label{eq:decompfield}
\Phi=\frac{1}{\sqrt{2}}(\varphi_b+\hat{\Phi}+\i G).
\end{align}
We consider the family of gauge-fixing terms of the form
\begin{align}
\mathcal{L}_{\rm G.F.}=\frac{{1}}{2\xi}(\partial_\mu\,A_\mu-\zeta\,g\,\varphi_b\,G)^2,
\end{align}
where $\zeta=0$~\citep{Endo:2017tsz,Andreassen:2017rzq} for Fermi gauge, $\zeta=1$~\citep{Isidori:2001bm} or $\zeta=\xi$~\cite{Branchina:2014rva} for $R_\xi$ gauge. 
For this family of gauges and to compute the one-loop effective action we consider the Lagrangian up to quadratic order in the field fluctuations, given that higher order interactions do not enter one-loop effects, thus:
\begin{align}
\label{eq:quadraticLagrangian}
\begin{aligned}
\mathcal{L}^{(2)}&=\frac{1}{2}\,(\partial_\mu\varphi_b)^2+\frac{\alpha}{2}\,\varphi_b^2+\,\frac{\lambda}{4}\,\varphi_b^4+\frac{\lambda_6}{8}\,\varphi_b^6 +\frac{1}{2}\,\hat{\Phi}\left(-\Box+\alpha+3\,\lambda\,\varphi_b^2+\frac{15\,\lambda_6}{4}\,\varphi_b^4\right)\,\hat{\Phi}\\
&\qquad + \frac{1}{2}\,A_\mu\,\left[(-\Box+g^2\,\varphi_b^2)\,\delta_{\mu\nu}+\frac{\xi-1}{\xi}\,\partial_\mu\partial_\nu\right]\,A_\nu\\
&\qquad + \frac{1}{2}\,G\,\left(-\Box+\alpha+\lambda\,\varphi_b^2+\frac{3\,\lambda_6}{4}\varphi_b^4+\frac{\zeta^2\,g^2\,\varphi_b^2}{\xi}\right)\,G\\
&\qquad +\bar{\eta}\,\left(-\Box+\zeta\,g^2\,\varphi_b^2\right)\,\eta
+\left(\frac{\zeta+\xi}{\xi}\right)\,g\,A_{\mu}\,(\partial_\mu\varphi_b)\,G +\left(\frac{\zeta-\xi}{\xi}\right)\,g\,\varphi_b\,A_\mu\,(\partial_\mu G),
\end{aligned}\end{align}
where we employ the notation $\Box$ to denote the four-dimensional Laplacian operator. 

The expansion in Eq.~\eqref{eq:quadraticLagrangian} allows to compute the partition function within a Gau{\ss}ian approximation. The current corresponding to the tree-level bounce is $J[\varphi_b]=\mathcal{O}(\hbar)$, as follows from a loop expansion of Eq.~\eqref{eq:currents}. Similarly, it can be seen that the one-point function associated with this current is $\varphi=\varphi_b+\mathcal{O}(\hbar).$ Taking this into account when calculating the effective action using Eq.~\eqref{eq:effectiveAction0} together with the Gau{\ss}ian approximation to the partition function, one obtains
\begin{align}
\label{eq3.13}\begin{aligned}
\Gamma^{(1)}[\varphi_b]-\Gamma^{(1)}[0]=&\,S[\varphi_b]+\frac{\hbar}{2}\,\log\frac{\det\mathcal{M}^{-1}_{\hat{\Phi}}(\varphi_b)}{\det\mathcal{M}^{-1}_{\hat{\Phi}}(0)}+\frac{\hbar}{2}\,\log\frac{\det\mathcal{M}^{-1}_{(A_\mu,G)}(\varphi_b)}{\det\mathcal{M}^{-1}_{(A_\mu,G)}(0)}\\
 &-\hbar\,\log\frac{\det\mathcal{M}^{-1}_{(\bar{\eta},\eta)}(\varphi_b)}{\det\mathcal{M}^{-1}_{(\bar{\eta},\eta)}(0)},
\end{aligned} \end{align} 
where 
\begin{align}
\label{eq:greenMatrices}
\begin{aligned}
&\mathcal{M}^{-1}_{\hat{\Phi}}(\varphi_b)=-\Box+\alpha+3\,\lambda\,\varphi_b^2+\frac{15\,\lambda_6}{4}\,\varphi_b^4\;,\\
&\mathcal{M}^{-1}_{(A_\mu,G)}(\varphi_b)=
\begin{pmatrix}
(-\Box+g^2\,\varphi_b^2)\,\delta_{\mu\nu}+\frac{\xi-1}{\xi}\partial_{\mu}\partial_{\nu} & \left(\frac{\zeta + \xi}{\xi}\right)g\,(\partial_\mu\varphi_b) + \left(\frac{\zeta-\xi}{\xi}\right)g\,\varphi_b\,\partial_\mu \\
2\,g\,(\partial_\nu\varphi_b)+\left(\frac{\xi-\zeta}{\xi}\right)g\,\varphi_b\,\partial_\nu & -\Box + \alpha + \lambda\,\varphi_b^2 + \frac{3\,\lambda_6}{4}\,\varphi_b^4 + \frac{\zeta^2\,g^2\,\varphi_b^2}{\xi} \\
\end{pmatrix}\;,\\
&\mathcal{M}^{-1}_{(\bar{\eta},\eta)}(\varphi_b)=-\Box+\zeta\,g^2\,\varphi_b^2\;.
\end{aligned}
\end{align}
Here $\mathcal{M}^{-1}_{(A_\mu,G)}$ is a $5\times 5$ matrix which operates on the vector $(A_\nu,G)^T$ from the left. In Eq.~\eqref{eq3.13} we have made use of the assumption that the tree-level potential is normalized to zero at the false vacuum $\varphi=0$. One can further choose counterterms such that the full effective action evaluated at $\varphi=0$ (equivalent to the effective potential at the origin) vanishes,  such that in the following we will set $\Gamma^{(1)}[0]=0$ in Eq.~\eqref{eq3.13}.  One can show that $\varphi^{(1)}=\varphi_b+\mathcal{O}(\hbar)$ via Eq.~\eqref{eq:quantumeom}. Using this, together with the aforementioned fact that  $J[\varphi_b]=\mathcal{O}(\hbar)$, we arrive at the one-loop effective action  evaluated at the one-loop, quantum-corrected bounce: 
\begin{align}
\label{eq:Gammaphi1}
\begin{array}{ll}
\Gamma^{(1)}[\varphi^{(1)}] &\displaystyle = S[\varphi^{(1)}] + \frac{\hbar}{2}\,\log\frac{\det\mathcal{M}^{-1}_{\hat{\Phi}}(\varphi^{(1)})}{\det\mathcal{M}^{-1}_{\hat{\Phi}}(0)}+\frac{\hbar}{2}\,\log\frac{\det\mathcal{M}^{-1}_{(A_\mu,G)}(\varphi^{(1)})}{\det\mathcal{M}^{-1}_{(A_\mu,G)}(0)}-\\[15pt]   &\qquad \displaystyle-\hbar\,\log\frac{\det\mathcal{M}^{-1}_{(\bar{\eta},\eta)}(\varphi^{(1)})}{\det\mathcal{M}^{-1}_{(\bar{\eta},\eta)}(0)}.\end{array}
\end{align}
The above expression and its higher-order generalization can be derived in a rigorous manner by tracking carefully the distinction between saddle points and one-point functions, and using the method of constrained sources to ensure that the saddle points of the relevant path integrals coincide with the quantum-corrected bounce instead of the tree-level bounce used in the previous derivation \cite{Garbrecht:2015cla}.  Eq.~\eqref{eq:Gammaphi1} will be our starting point, and we will do a perturbative expansion of the quantum bounce $\varphi^{(1)}$ around an initial approximation $\varphi_0$---to be referred to as the ``simplified bounce''---which may not necessarily be the tree-level bounce $\varphi_b$, but e.g. the bounce computed using the one-loop Coleman-Weinberg potential in which possible gradient corrections have been neglected.

Before implementing this expansion, we introduce a further simplification making use of the planar-wall approximation. In the limit in which the false and true vacua become degenerate, the radius of the bubble described by the bounce configuration grows, and it can become so large that the bubble can be approximated by a planar configuration. This allows us to trade the determinants of operators over functions in $\mathbb{R}^4$ in Eq.~\eqref{eq:Gammaphi1} with simpler determinants over momentum-dependent operators acting on functions in $\mathbb{R}$. In this approximation, the bubble profile depends on a single Cartesian coordinate (corresponding to the direction perpendicular to the wall) that we choose to be $x_4$. We may further shift $x_4$ to a coordinate $z$, such that $z=0$ at the center the bubble wall, defined by the location of its steepest gradient. In this approximation the fluctuation operators  ${\cal M}^{-1}_X(\varphi^{(1)})$---with $X\in\{\hat\Phi,(A_\mu,G),(\bar\eta,\eta)\}$---become independent of ${\bf x}=\{x_1,x_2,x_3\}$, and thus one can consider a basis of eigenfunctions of the form
\begin{align}
\label{eq:planareigenfunctions}
 \phi_{X;{\bf k},i}(x)=\frac{\e^{\i{\bf k}\cdot{\bf x}}}{(2\pi)^{3/2}}\,f_{X;{\bf k},i}(z) \quad \text{with} \quad {\cal M}^{-1}_X(\varphi^{(1)})\phi_{X;{\bf k},i}=\lambda_{X;{\bf k},i}\,\phi_{X;{\bf k},i}.
\end{align}
Inserting the definitions of the fluctuation operators in Eq.~\eqref{eq:greenMatrices}, one ends up with eigenvalue equations for the functions $f_{X;{\bf k},i}(z)$ involving planar fluctuation operators ${\cal M}^{-1}_{X;{\bf k}}(\varphi^{(1)})$:
\begin{align}
\label{eq:planareigenvalueeq}
 {\cal M}^{-1}_{X;{\bf k}}(\varphi^{(1)}) f_{X;{\bf k},i}=  \lambda_{X;{\bf k},i}\,f_{X;{\bf k},i},
\end{align}
with
\begin{align}
\label{eq:koperators}\begin{aligned}
\mathcal{M}^{-1}_{\hat{\Phi};{\bf k}}(\varphi^{(1)})=&
-\partial_z^2+{\bf k}^2+\alpha+3\,\lambda\,{\varphi^{(1)2}}+\frac{15\,\lambda_6}{4}\,{\varphi^{(1)4}},\\
\mathcal{M}^{-1}_{(\bar{\eta},\eta);{\bf k}}(\varphi^{(1)})=&\,-\partial_z^2+{\bf k}^2+\zeta\,g^2\,{\varphi^{(1)2}},\\
\mathcal{M}^{-1}_{(A_\mu,G);{\bf k}}(\varphi^{(1)})=\\
&\hskip-2cm\begin{pmatrix}
M_{\bf k}^{-1}(\varphi^{(1)})\delta_{ij}+\frac{\xi-1}{\xi}(\i\mathbf{k}_i)(\i\mathbf{k}_j) & \frac{\xi-1}{\xi}(\i\mathbf{k}_i)\partial_z & \frac{\zeta-\xi}{\xi}g\varphi^{(1)}(\i\mathbf{k}_i) \\
\frac{\xi-1}{\xi}(\i\mathbf{k}_j)\partial_z & M_{\bf k}^{-1}(\varphi^{(1)})+\frac{\xi-1}{\xi}\partial_z^2 & \left(\frac{\zeta+\xi}{\xi}\right)g(\partial_z\varphi^{(1)})+\frac{\zeta-\xi}{\xi}g\varphi^{(1)}\partial_z \\
\left(\frac{\xi-\zeta}{\xi}\right)g\varphi^{(1)}(\i\mathbf{k}_j) & 2g(\partial_z\varphi^{(1)})+\left(\frac{\xi-\zeta}{\xi}\right)g\varphi^{(1)}\partial_z & N_{\bf k}^{-1}(\varphi^{(1)}) \\
\end{pmatrix}.
\end{aligned}\end{align}
Here
\begin{align}\begin{aligned}
M_{{\bf k}}^{-1}(\varphi^{(1)})&=-\partial_z^2+{\bf k}^2+g^2\,{\varphi^{(1)2}},\\
N_{{\bf k}}^{-1}(\varphi^{(1)})&=-\partial_z^2+{\bf k}^2+\alpha+\lambda\,{\varphi^{(1)2}}+\frac{3\,\lambda_6}{4}{\varphi^{(1)4}}+\frac{\zeta^2\,g^2\,{\varphi^{(1)2}}}{\xi}.
\end{aligned}\end{align}
Observe that when choosing $\xi=\zeta=1$, the fields $A_i\  (i=1,2,3)$ neatly decouple and the planar fluctuation operators for $A_i$ are the same as that for ghost fields, leading to a cancellation of the contribution from the ghost fields with one of the three gauge field degrees of freedom. 

Due to the Hermiticity of the operators ${\cal M}^{-1}_{X;{\bf k}}(\varphi^{(1)})$, the eigenfunctions $f_{X;{\bf k},i}(z)$ are orthogonal, and so are the corresponding eigenfunctions $\phi_{X;{\bf k},i}(x)$ of the full problem.  The eigenfunctions $f_{X;{\bf k},i}(z)$ are assumed to have the usual normalization for either the discrete spectrum (with the scalar product of eigenfunctions with indices $i$ and $j$ being a Kronecker delta $\delta_{ij}$) or the continuum spectrum (the scalar product being a delta function $\delta(i-j)$). Given this, one can write a spectral decomposition of  $\log {\cal M}^{-1}_{X}(\varphi^{(1)})$ into orthogonal projectors. Representing operators $\cal O$ in terms of matrices ${\cal O}(x,y)$ with continuous indices, such that $({\cal O }f)(x) =\int {\cal O}(x,y)f(y)$, one gets:
\begin{align}\label{eq:logMk0}\begin{aligned}
 \log{\cal M}^{-1}_{X}(\varphi^{(1)};x',x)=&\,\int {\d^3{\bf k}}\SumInt_{i}\log(\lambda_{X;{\bf k},i}) \phi_{X;{\bf k},i}(x')\phi^\dagger_{X;{\bf k},i}(x)\\ 
 =&\,\int \frac{\d^3{\bf k}}{(2\pi)^3}\e^{\i{\bf k}\cdot({\bf x'}-{\bf x})}\SumInt_{i}\log(\lambda_{X;{\bf k},i}) P_{X;{\bf k},i}(z',z).
\end{aligned}\end{align}
In the previous equation,  $\SumInt$ denotes a sum over discrete values and integration over continuum values. The operators $P_{X;{\bf k},i}(z',z)=f_{X;{\bf k},i}(z')f^\dagger_{X;{\bf k},i}(z)$ are projectors onto the eigenfunctions of the planar fluctuation operators ${\cal M}^{-1}_{X;{\bf k}}(\varphi^{(1)})$ of Eqs.~\eqref{eq:planareigenvalueeq}, \eqref{eq:koperators}. Then $\SumInt_i\log \lambda_{X;{\bf k},i} P_{X;{\bf k},i}$ is nothing but the spectral decomposition of the operator $\log {\cal M}^{-1}_{X;{\bf k}}(\varphi^{(1)})$, and thus one ends up with
\begin{align}\label{eq:logMk}\begin{aligned}
 \log{\cal M}^{-1}_{X}(\varphi^{(1)};x',x)=\int \frac{\d^3{\bf k}}{(2\pi)^3}\e^{\i{\bf k}\cdot({\bf x'}-{\bf x})}\log{\cal {\cal M}}^{-1}_{X;{\bf k}}(\varphi^{(1)};z',z).
\end{aligned}\end{align}
In the same manner one can show that the Green's functions ${\cal M}_{X}(\varphi^{(1)})$ (the inverses of the planar fluctuation operators ${\cal M}^{-1}_{X}(\varphi^{(1)})$), satisfy
\begin{align}\label{eq:Mk}\begin{aligned}
{\cal M}_{X}(\varphi^{(1)};x',x)=\int \frac{\d^3{\bf k}}{(2\pi)^3}\e^{\i{\bf k}\cdot({\bf x'}-{\bf x})}{\cal M}_{X;{\bf k}}(\varphi^{(1)};z',z).
\end{aligned}\end{align}
One can then compute $\log\det  {\cal M}^{-1}_{X}={\rm Tr}\log  {\cal M}^{-1}_{X}$ (with ${\rm Tr}$ acting on continuous as well as discrete indices) as follows:
\begin{align}\begin{aligned}
 \log\det  {\cal M}^{-1}_{X}(\varphi^{(1)})= &\tr\,\int \d^4x\log{\cal M}^{-1}_{X}(\varphi^{(1)};x,x)=V{\rm Tr}_z\int \frac{\d^3{\bf k}}{(2\pi)^3}\log{\cal M}^{-1}_{X;{\bf k}}(\varphi^{(1)})\\
 =&\,V\int \frac{\d^3{\bf k}}{(2\pi)^3}\log \det {\cal M}^{-1}_{X;{\bf k}},
\end{aligned}\end{align}
where ``$\tr$'' denotes a trace over the remaining discrete matrix structure, if any (e.g. for $X=(A_\mu,G)$, see Eq.~\eqref{eq:koperators}).
From this result, we can finally obtain an expression for the effective action evaluated at the quantum bounce,  Eq.~\eqref{eq:Gammaphi1}, in the planar-wall approximation:
\begin{align}
\label{eq:finaleffectiveaction}
\Gamma^{(1)}[\varphi^{(1)}]&=S[\varphi^{(1)}]+\frac{\hbar}{2}\,V\int\frac{\D^3{\bf k}}{(2\pi)^3}\,\log\frac{\det\mathcal{M}^{-1}_{\hat{\Phi};{\bf k}}(\varphi^{(1)})}{\det\mathcal{M}^{-1}_{\hat{\Phi};{\bf k}}(0)}+\frac{\hbar}{2}\,V\tr\int\frac{\D^3{\bf k}}{(2\pi)^3}\log\frac{\det\mathcal{M}^{-1}_{(A_\mu,G);{\bf k}}(\varphi^{(1)})}{\det\mathcal{M}^{-1}_{(A_\mu,G);{\bf k}}(0)}\notag\\
&-\hbar\,V\int\frac{\D^3{\bf k}}{(2\pi)^3}\log\frac{\det\mathcal{M}^{-1}_{(\bar{\eta},\eta);{\bf k}}(\varphi^{(1)})}{\det\mathcal{M}^{-1}_{(\bar{\eta},\eta);{\bf k}}(0)}.
\end{align}
To end this section, let us comment on the interpretation of $\Gamma^{(1)}[\varphi^{(1)}]$ in terms of Feynman diagrams. The functional $\Gamma^{(1)}[\varphi^{(1)}]$ includes additional two-loop corrections with respect to the one-loop  effective action evaluated at the initial estimate of the bounce $\varphi_0$, $\Gamma^{(1)}[\varphi_0]$. When $\varphi_0$ is the tree-level bounce and 
when the propagators are understood as Green's functions in the background
of $\varphi_0$, these corrections to the effective action correspond to dumbbell diagrams, as shown in Figures~\ref{fig:tadpoles} and~\ref{fig:dumbbell}~\cite{Garbrecht:2015oea}. While these corrections are of two-loop order, they may be enhanced over additional two-loop effects. For example, in non-Abelian theories the summation over colour enhances the dumbbells with respect to other topologies such as sunset diagrams. Note that the fact that the two-loop diagrams contributing to $\Gamma^{(1)}[\varphi^{(1)}]$ are not one-particle irreducible is not contradicting the usual properties of the effective action. This is just an artifact of defining propagators in the background of the simplified bounce $\varphi_0$, rather than $\varphi^{(1)}$. Doing the latter, the diagrams that contribute to $\Gamma[\varphi^{(1)}]$ are always one-particle irreducible \cite{Garbrecht:2015oea}.

\subsection{Green's function method}
\label{subsec:Green}

Up to now,  $\varphi^{(1)}$ is yet undetermined. In this section, we derive its equation of motion. For this purpose, we expand Eq.~\eqref{eq:finaleffectiveaction} around $\varphi_0$. Writing $\varphi^{(1)}=\varphi_0+\hbar\delta\varphi$, one obtains 
{\begin{align}
\label{eq:effectiveAction1}
\Gamma^{(1)}[\varphi^{(1)}]\ &=\ B+\hbar\, B^{(1)} +\hbar\, B^{(1)}_{\hat{\Phi};{\rm dis}}+\hbar\, B_{\hat{\Phi}}^{(1)}+\hbar\, B_{(A_\mu,
G)}^{(1)}+\hbar\, B_{(\bar{\eta},\eta
)}^{(1)}+\hbar^2\, B^{(2)}+\hbar^2\,B_{\hat{\Phi}}^{(2)}\notag \\[6pt]
 & +\hbar^2\,B_{(A_\mu,
G)}^{(2)}+\hbar^2\,B_{(\bar{\eta}\;,
\eta)}^{(2)},
\end{align}
 where the different contributions are explained in the following. First, we recall that $B$ is the classical action evaluated at the simplified bounce, $B=S[\varphi_0]$. $B^{(1)}$ and $B^{(2)}$ are related to the expansion of the classical action around $\varphi_0$, while the rest of the terms originate from the expansion of  the one-loop corrections. Starting with the latter, the first term
 }
\begin{align}
\label{eq:discrete}
\hbar B^{(1)}_{\hat{\Phi};{\rm dis}}=\frac{\i\pi\hbar}{2} -\frac{\hbar}{2}\log\left(\frac{(V{\cal T})^2\alpha^5}{{4}|\lambda_0|}\left(\frac{B}{2\pi\hbar}\right)^4\right)
\end{align}
is the contribution from the discrete modes to the scalar fluctuations. Four zero modes corresponding spacetime translations of the scalar field fluctuations have been traded for collective coordinates. In addition, there is one negative
mode about the bounce, which   in the thin-wall approximation takes the value~\cite{Garbrecht:2015oea,Garbrecht:2018rqx}
\begin{align}
\lambda_0=-\frac{3}{R^2},
\end{align}
where $R$ is the radius of the critical bubble. {The integral over this negative mode needs careful analytic continuation which leads to half the result obtained when one naively performs the Gaussian integral as if $\lambda_0$ is positive, which explains both the term $\i\pi\hbar/2$ and the factor of $4$ inside the logarithm in Eq.~\eqref{eq:discrete}~\cite{Callan:1977pt}.} In the planar-wall approximation, these five discrete modes appear for $\mathbf k=0$. Because of the vanishing integration measure at this point, we pair these up explicitly with the lowest eigenvalues $\alpha$ at the bottom of the continuum spectrum about the false vacuum for $\mathbf{k}=0$. This is to be compared with the example of the one-dimensional kink, where no planar-wall approximation is made and the dimensionless factors
in the determinant quotients match up without ado~\cite{Ai:2019fri}. 

The remaining $\hbar B^{(1)}_X$ contributions are expressed in terms of functional determinants as
\begin{subequations}
\label{eq:B1}
\begin{align}
&{ \hbar}B_{\hat{\Phi}}^{(1)}=
\frac{ \hbar}{2}\,V\int\frac{\D^3{\bf k}}{(2\pi)^3}\,\log\left|\frac{\det\mathcal{M}^{-1}_{\hat{\Phi};{\bf k}}(\varphi_0)}{\det\mathcal{M}^{-1}_{\hat{\Phi};{\bf k}}(0)}\right|,\\
&{ \hbar}B_{(A_\mu,G)}^{(1)}=\frac{ \hbar}{2}\,V\tr\int\frac{\D^3{\bf k}}{(2\pi)^3}\log\frac{\det\mathcal{M}^{-1}_{(A_\mu,G);{\bf k}}(\varphi_0)}{\det\mathcal{M}^{-1}_{(A_\mu,G);{\bf k}}(0)},\\
&{ \hbar}B_{(\bar{\eta},\eta)}^{(1)}=-{ \hbar}\,V\int\frac{\D^3{\bf k}}{(2\pi)^3}\log\frac{\det\mathcal{M}^{-1}_{(\bar{\eta},\eta);{\bf k}}(\varphi_0)}{\det\mathcal{M}^{-1}_{(\bar{\eta},\eta);{\bf k}}(0)}.
\end{align}
\label{eq:determinantContrib}
\end{subequations}
Above, the trace ``$\tr$'' is again  understood to apply to the matrix structure of the operators. Note that the discrete modes yield a vanishing contribution to the ${\rm d}^3 \mathbf{k}$ integration in the planar-wall approximation
and therefore do not need to be dealt with explicitly in the above expression for
$B^{(1)}_\Phi$.

To obtain the functional determinants, we follow  e.g. Refs.~\cite{Baacke:1993aj,Baacke:1993jr,Baacke:2008zx,Ai:2018guc} and generalize the Green's functions ${\cal M}_{X;\mathbf{k}}(\varphi;z,z^\prime)$ to resolvents 
\begin{align}
 {\cal M}_{X;\sqrt{\mathbf{k}^2+s}}(\varphi;z,z^\prime)=\left.{\cal M}_{X;{\mathbf{k}}}(\varphi;z,z^\prime)\right|_{{\bf k}^2\rightarrow{\bf k}^2+s}.
\end{align}
The logarithm of the ratio of functional determinants is then
given by
\begin{align}
\label{eq:functionalDetDeformation}
\begin{aligned}
\log\frac{\det \mathcal{M}^{-1}_X(\varphi)}{\det\mathcal{M}^{-1}_X(0)} &=V\tr\int\frac{\D^3{\bf k}}{(2\pi)^3}\log \frac{\det \mathcal{M}^{-1}_{X;{\bf k}}(\varphi)}{\det\mathcal{M}^{-1}_{X;{\bf k}}(0)}\\
&\hskip-1.5cm=-\tr\int_{-\infty}^\infty \d z\int \d^3 {\bf x} \int_0^\infty \d s \int\frac{\D^3{\bf k}}{(2\pi)^3}
\left(\mathcal{M}_{X;\sqrt{{\bf k}^2+s}}(\varphi;z,z)-\mathcal{M}_{X;\sqrt{{\bf k}^2+s}}(0;z,z)\right).
\end{aligned}\end{align}
The resolvent is a generalization of the Green's
function that is most straightforwardly defined through a spectral sum over the eigenmodes as e.g. in Refs.~\cite{Garbrecht:2015yza,Ai:2019fri}. Green's functions and resolvents for scalar fields in the background of tunnelling solutions have been found in Refs.~\cite{Garbrecht:2015oea,Garbrecht:2015yza}. In the present work, we focus on constructing the Green's function in the gauge-Goldstone sector, where the details are explained in Section~\ref{sec:solvingA4G}.

At next order in the expansion in $\hbar$ of $\Gamma^{(1)}[\varphi^{(1)}]$, there are terms of the form
\begin{align}
\label{B2:dumbbell}
\begin{aligned}
&{ \hbar^2}B_{\hat{\Phi}}^{(2)}=\frac{ \hbar}{2}\,V\int\frac{\D^3{\bf k}}{(2\pi)^3}\,\int\D z\,{ \hbar}\delta\varphi(z)\,\frac{\delta}{\delta\varphi^{}(z)}\log\left.\frac{\det\mathcal{M}^{-1}_{\hat{\Phi};{\bf k}}(\varphi^{})}{\det\mathcal{M}^{-1}_{\hat{\Phi};{\bf k}}(0)}\right|_{\varphi_0},\\
&{ \hbar^2}B_{(A_\mu,G)}^{(2)}=\frac{ \hbar}{2}\,V\int\frac{\D^3{\bf k}}{(2\pi)^3}\int\D z\,{ \hbar}\delta\varphi(z)\,\frac{\delta}{\delta\varphi^{}(z)}\left.\log\frac{\det\mathcal{M}^{-1}_{(A_\mu,G);{\bf k}}(\varphi^{})}{\det\mathcal{M}^{-1}_{(A_\mu,G);{\bf k}}(0)}\right|_{\varphi_0},\\
&{ \hbar^2}B_{(\bar{\eta},\eta)}^{(2)}=-{\hbar}\,V\int\frac{\D^3{\bf k}}{(2\pi)^3}\int \D z\,{ \hbar}\delta\varphi(z)\,\frac{\delta}{\delta\varphi^{}(z)}\,\left.\log\frac{\det\mathcal{M}^{-1}_{(\bar{\eta},\eta);{\bf k}}(\varphi^{})}{\det\mathcal{M}^{-1}_{(\bar{\eta},\eta);{\bf k}}(0)}\right|_{\varphi_0}.
\end{aligned}
\end{align}
As anticipated earlier and as will be justified below, these correspond to two-loop dumbbell diagrams. Additional dumbbell contributions arise when expanding the classical action to second order in $\hbar\delta\varphi$, which gives rise to the $B^{(2)}$ contribution in Eq.~\eqref{eq:effectiveAction1}. In order to extract these terms, we functionally differentiate Eq.~\eqref{eq:finaleffectiveaction} with respect to $\varphi^{(1)}$. This yields the equation of motion for the corrected bounce,
\begin{align}
-\partial_z^2\varphi^{(1)}(z)+U'_{\rm eff}(\varphi^{(1)};z)=0,
\label{eq:EoMbounce1}
\end{align}
where 
\begin{align}
U'_{\rm eff}(\varphi^{(1)};z)&=U'(\varphi^{(1)};z)+\hbar\,\Pi_
{\hat{\Phi}}(\varphi_0;z)\,\varphi_0(z)+\hbar\,\Pi_
{(A_\mu,G)}(\varphi_0;z)\,\varphi_0(z)\notag\\
&+\hbar\,\Pi_
{(\bar{\eta},\eta)}(\varphi_0;z)\,\varphi_0(z)
\end{align}
and
\begin{align}\label{eq:Vphi1}
U(\varphi^{(1)})=\frac{1}{2}\alpha\varphi^{(1)2}+\frac{\lambda}{4}\varphi^{(1)4}+\frac{\lambda_6}{8}\varphi^{(1)6}.
\end{align}
The functions $\Pi_{\hat{\Phi}}$, $\Pi_{(A_\mu,G)}$ and $\Pi_{(\bar{\eta},\eta)}$ may be interpreted as self-energy contributions from the field fluctuations, which are derived from the functional derivatives of the various contributions to $B^{(1)}$ with respect to the background field, giving the so-called tadpole terms:
\begin{subequations}
 \label{eq:Piphiall}
\begin{align}
{\hbar}\,\Pi_{\hat{\Phi}}(\varphi_0;z)\,\varphi_0(z)&=\frac{\hbar}{2}\,\int\frac{\D^3{\bf k}}{(2\pi)^3}[6\,\lambda\,\varphi_0(z)+15\,\lambda_6\,\varphi_0^3(z)]\,\mathcal{M}_{\hat{\Phi};{\bf k}}(\varphi_0;z,z),\label{eq:PiPhi}\\
{\hbar}\,\Pi_{(A_\mu,G)}(\varphi_0;z)\,\varphi_0(z)&=\frac{{\hbar}}{2}\,\int\frac{\D^3{\bf k}}{(2\pi)^3}\, \tr \left(\mathcal{M}_{(A_\mu,G);{\bf k}}(\varphi_0;z,z)\left.\frac{\partial
	\mathcal{M}^{-1}_{(A_\mu,G);{\bf k}}(\varphi)}{\partial\varphi}(z)\right|_{\varphi_0}\right),\label{eq:PiA4G}\\
{\hbar}\,\Pi_{(\bar{\eta},\eta)}(\varphi_0;z)\,\varphi_0(z) &=-2\,{\hbar}\,\zeta\,g^2\,\int\frac{\D^3{\bf k}}{(2\pi)^3}\, \mathcal{M}_{(\bar{\eta},\eta);{\bf k}}(\varphi_0;z,z)\,{\varphi_0}(z).\label{eq:PiEta}
\end{align}
\end{subequations}

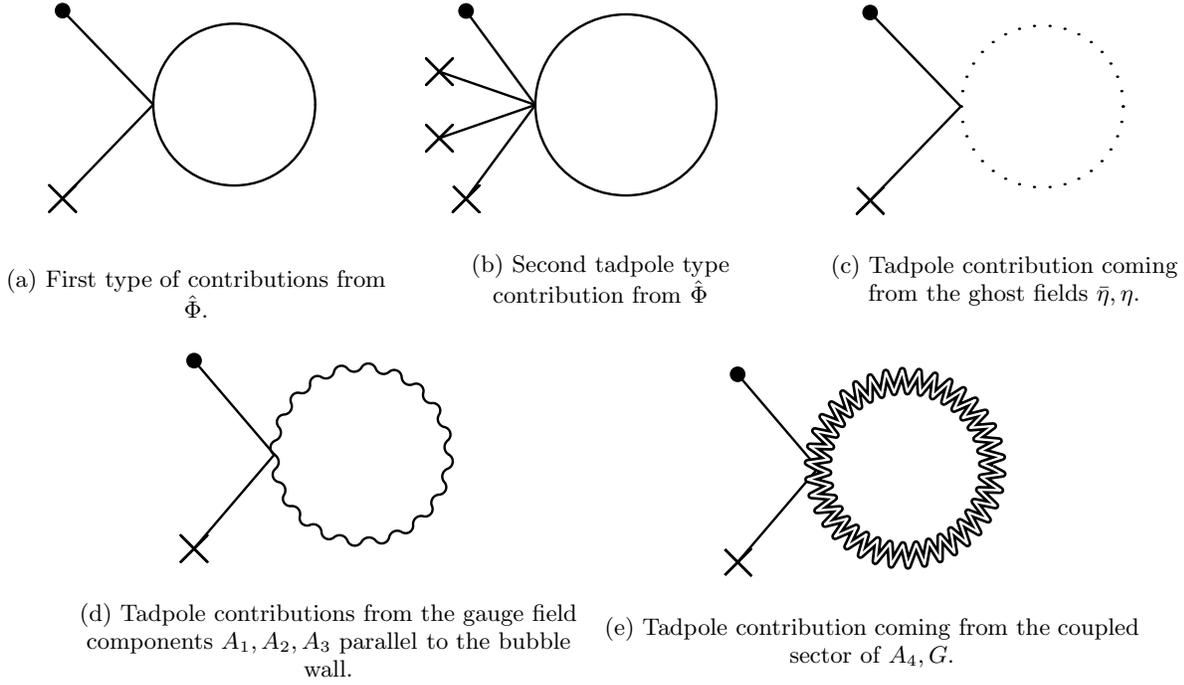
\begin{figure}
	\centering
	\begin{subfigure}{0.32\textwidth}
	\begin{fmffile}{./loop1}
		\fmfframe(0,2)(0,2){ %
		\begin{fmfgraph*}(40,25)
					\fmfleft{i1,i2}
					\fmfright{o}
					\fmfv{decor.shape=cross}{i1}
					\fmfv{decor.shape=circle,decor.filled=full,decor.size=5}{i2}
					\fmf{plain,tension=5}{i1,v1}
					\fmf{plain,tension=5}{i2,v1}
					\fmf{plain,left,tension=0.3}{v1,v2,v1}
					\fmf{phantom,tension=50}{v2,o}
					\fmf{phantom,tension=5}{v1,v2}
		\end{fmfgraph*}
			}
	\end{fmffile}
		\vspace*{5mm}
		\caption{First type of contributions from $\hat{\Phi}$.}
		\label{fig:loopScalar}
	\end{subfigure}
	\hfill
	\begin{subfigure}{0.32\textwidth}
		\begin{fmffile}{./loop2}
				\begin{fmfgraph}(40,25)
					\fmfleft{i1,i2,i3,i4}
					\fmfright{o}
					\fmfv{decor.shape=cross}{i1,i2,i3}
					\fmfv{decor.shape=circle,decor.filled=full,decor.size=5}{i4}
					\fmf{plain,tension=.4}{i1,v1}
					\fmf{plain,tension=5}{i2,v1}
					\fmf{plain,tension=5}{i3,v1}
					\fmf{plain,tension=.4}{i4,v1}
					\fmf{plain,left,tension=0.3}{v1,v2,v1}
					\fmf{phantom,tension=50}{v2,o}
					\fmf{phantom,tension=5}{v1,v2}
				\end{fmfgraph}
			\end{fmffile}
		\vspace*{5mm}
		\caption{Second tadpole type contribution from $\hat{\Phi}$}
		\label{fig:loopScalar2}
	\end{subfigure}
	\hfill
	\begin{subfigure}{0.32\textwidth}
	\begin{fmffile}{./loop3}
			\begin{fmfgraph}(40,25)
				\fmfleft{i1,i2}
				\fmfright{o}
				\fmfv{decor.shape=cross}{i1}
				\fmfv{decor.shape=circle,decor.filled=full,
				decor.size=5}{i2}
				\fmf{plain,tension=5}{i1,v1}
				\fmf{plain,tension=5}{i2,v1}
				\fmf{dots,left,tension=0.3}{v1,v2,v1}
				\fmf{phantom,tension=50}{v2,o}
				\fmf{phantom,tension=5}{v1,v2}
			\end{fmfgraph}
		\end{fmffile}
		\vspace*{5mm}
		\caption{Tadpole contribution coming from the ghost fields $\bar{\eta},\eta$.}
		\label{fig:loopGauge}
	\end{subfigure}\\[10pt]
	\begin{subfigure}{0.45\textwidth}
		\begin{fmffile}{./loop4}
			\begin{fmfgraph}(40,25)
				\fmfleft{i1,i2}
				\fmfright{o}
				\fmfv{decor.shape=cross}{i1}
				\fmfv{decor.shape=circle,decor.filled=full,						decor.size=5}{i2}
				\fmf{plain,tension=5}{i1,v1}
				\fmf{plain,tension=5}{i2,v1}
				\fmf{phantom,tension=50}{v2,o}
				\fmf{wiggly,left,tension=0.3}{v1,v2,v1}
				\fmf{phantom,tension=4}{v1,v2}
			\end{fmfgraph}
		\end{fmffile}
		\vspace*{5mm}
		\caption{Tadpole contributions from the gauge field components $A_1,A_2,A_3$ parallel to the bubble wall.}
		\label{fig:loopGaugeParallel}
	\end{subfigure}
	\begin{subfigure}{0.45\textwidth}
		\begin{fmffile}{./loop5}
					\begin{fmfgraph}(40,25)
						\fmfleft{i1,i2}
						\fmfright{o}
						\fmfv{decor.shape=cross}{i1}
						\fmfv{decor.shape=circle,decor.filled=full,
										decor.size=5}{i2}
						\fmf{plain,tension=5}{i1,v1}
						\fmf{plain,tension=5}{i2,v1}
						\fmf{phantom,tension=50}{v2,o}
						\fmf{dbl_zigzag,left,tension=0.3}{v1,v2,v1}
						\fmf{phantom,tension=4}{v1,v2}
					\end{fmfgraph}
				\end{fmffile}
		\vspace*{5mm}
		\caption{Tadpole contribution coming from the coupled sector of $A_4,G$.}
		\label{fig:loopA4G}
	\end{subfigure}
	\captionsetup{justification=raggedright,singlelinecheck=false}
	\caption{Diagrammatic representation (with propagators defined in the background of the simplified bounce $\varphi_0$) of the tadpole terms associated with the functional derivatives of the one-loop contributions $B^{(1)}_X$ to the effective action. Solid lines correspond to the real scalar field $\hat\Phi$, dotted lines to ghosts, single wavy lines to the gauge field components parallel to the wall, and double wavy lines to the fluctuations in the mixed $A_4/G$ sector. The scalar lines ending in crosses represent  insertions of the background $\varphi_0$, while the lines ending in dots correspond to fluctuations around the background.}
	\label{fig:tadpoles} 
\end{figure}

Note that the tadpoles correspond to loop integrals of Green's functions ${\cal M}_{X;{\bf k}}(\varphi_0)$ in the background of the simplified bounce, multiplied by powers of the background field. Thus they correspond to amputations of the tadpole diagrams of Figure~\ref{fig:tadpoles}.

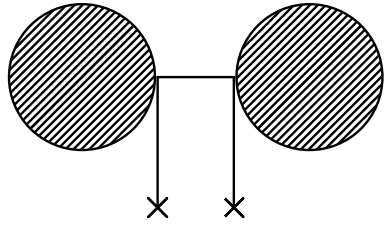
\begin{figure}
	\centering
		\begin{fmffile}{./dumbbell}
			\begin{fmfgraph}(50,35)
			\fmftop{t0,t1,t2,t3,t4,t5}
			\fmfbottom{b0,b1,b2,b3,b4,b5}
			\fmf{phantom}{t1,v1,b1}
			\fmf{phantom}{t2,v2,b2}
			\fmf{phantom}{t3,v3,b3}
			\fmf{phantom}{t4,v4,b4}
			\fmffreeze
			\fmfblob{55}{v1,v4}
			\fmf{plain}{v2,v3}
			\fmf{plain}{b2,v2}
			\fmf{plain}{b3,v3}
			\fmfv{decor.shape=cross,decor.size=10}{b2,b3}
			\end{fmfgraph}
		\end{fmffile}
		\vspace*{5mm}
	\captionsetup{justification=raggedright,singlelinecheck=false}
	\caption{\label{fig:dumbbell} Dumbbells diagrams (with propagators in the $\varphi_0$ background) corresponding to the two-loop corrections $B^{(2)}_X$ to the effective action. For simplicity we only represent one fluctuation leg and one background leg per vertex; the shaded blobs can be completed so as to match any of the tadpole diagrams in  Figure~\ref{fig:tadpoles}. }
\end{figure}

From Eqs.~\eqref{B2:dumbbell} and~\eqref{eq:Piphiall} it can be easily seen that
\begin{align}
\label{eq:B2Piphi}
 B^{(2)}_X=V\int \d z\,\delta\varphi(z)\,\Pi_X(\varphi_0;z) \varphi_0(z).
\end{align}
Substituting $\varphi^{(1)}=\varphi_0+\hbar\delta\varphi$ into Eq.~\eqref{eq:EoMbounce1}, one finds the following equation for $\delta\varphi(z)$:
\begin{align}
\label{eq:Mdeltavarphi}
\mathcal{M}^{-1}_{\hat{\Phi}}(\varphi_0;z)\,\delta\varphi(z)=&\,\frac{1}{\hbar}\left(\Box\varphi_0-U'(\varphi_0;z)\right)-\,\Pi_
{\hat{\Phi}}(\varphi_0;z)\,\varphi_0(z)-\,\Pi_
{(A_\mu,G)}(\varphi_0;z)\,\varphi_0(z)\notag\\
&-\,\Pi_
{(\bar{\eta},\eta)}(\varphi_0;z)\,\varphi_0(z).
\end{align}
 
The first contribution in the r.h.s of Eq.~\eqref{eq:Mdeltavarphi} vanishes if $\varphi_0$ is chosen as the classical bounce. Since as discussed above the $\Pi_X(\varphi_0;z)\varphi_b(z)$ correspond to amputated tadpole diagrams, Eq.~\eqref{eq:Mdeltavarphi} relates $\delta\varphi(z)$ to the full tadpole diagrams of Figure~\ref{fig:tadpoles}, as follows from the extra insertion of the propagator $\mathcal{M}_{\hat{\Phi}}(\varphi_0)$. Then,  Eq.~\eqref{eq:B2Piphi} implies  that, as mentioned earlier, the $B^{(2)}$ corrections are given by the dumbbell diagrams of Figure~\ref{fig:dumbbell}.
Finally, the remaining terms correspond to the expansion of the classical action,
\begin{align}
S[\varphi^{(1)}]=B+\hbar B^{(1)}+\hbar^2\, {B^{(2)}}.
\end{align}
The second term on the right hand side is
\begin{align}\label{eq:B10}
  B^{(1)}=V\int \d z \,\delta\varphi(z)\left.\frac{\delta S[\varphi]}{\delta\varphi(z)}\right|_{\varphi_0}=V\int \d z \,\delta\varphi(z)(-\Box\varphi_0+U'(\varphi_0;z)),
 \end{align}
which vanishes if $\varphi_0$ is taken as the classical bounce. On the other hand,
 using Eq.~\eqref{eq:Mdeltavarphi} together with \eqref{eq:B10} and \eqref{eq:B2Piphi} one obtains
\begin{align}
\label{eq:B2}
{B^{(2)}}&=\frac{1}{2}\,
\int\D^4 x\,\delta\varphi(z)\,\mathcal{M}^{-1}_{\hat{\Phi}}(\varphi_0;z)\,\delta\varphi(z)+\mathcal{O}(\hbar^2)\notag\\
&=-\frac{1}{2}\,\left(\,B_{\hat{\Phi}}^{(2)}+\,B_{(A_\mu,
G)}^{(2)}+\,B_{(\bar{\eta},\eta)}^{(2)}\right)-\frac{1}{2\hbar}B^{(1)}.
\end{align}
We see that that these contributions can be added to those of Eqs.~\eqref{eq:B10} and \eqref{B2:dumbbell}; when $\varphi_0$ is taken as the classical bounce, with $B^{(1)}=0$,  these terms are  again of the dumbbell type.

Collecting all these results, we obtain the tunnelling rate per unit volume:
\begin{align}
\label{eq:expandedDecayRate}
\begin{aligned}
\frac{\gamma}{V}=&\,\left(\frac{B}{2\pi\hbar}\right)^2{\sqrt\frac{\alpha^5}{{}{|\lambda_0|}}}\,\exp\left[-\frac{1}{\hbar}\,\left({\cal B}^{(0)}+\hbar\,{\cal B}^{(1)}+\hbar^2\,{\cal B}^{(2)}\right)\right],\\
{\cal B}^{(0)}=&\,B,\\
{\cal B}^{(1)}=&\,B_{\hat{\Phi}}^{(1)}+
B_{(A_\mu,G)}^{(1)}+B_{(\bar{\eta},
\eta)}^{(1)},\\
{\cal B}^{(2)}=&\,-{B^{(2)}}.
\end{aligned}\end{align}

Having organized the expansion in this form we see that, in order to include gradient effects in the calculation of the decay rate of the false vacuum, the main task that needs to be carried out is to compute the coincident limit $z'\rightarrow z$ of the ${\bf k}$-dependent Green's functions ${\cal M}_{X;{\bf k}}(\varphi_0;z,z')$ and resolvents ${\cal M}_{X;{\sqrt{\bf k^2+s}}}(\varphi_0;z,z')$ in the background of the {\it simplified} bounce $\varphi_0$. From the coincident Green's functions one can readily obtain the $B^{(1)}_X$ corrections from Eqs.~\eqref{eq:B1} and~\eqref{eq:functionalDetDeformation}. The  tadpoles $\Pi_X (\varphi_0;z)\varphi_0(z)$ can be obtained using Eqs.~\eqref{eq:Piphiall}, which then fixes the $B^{(2)}$ correction by means of Eqs.\eqref{eq:B2}, \eqref{eq:B10}, \eqref{eq:B2Piphi} and~\eqref{eq:Mdeltavarphi}.


\section{Solving for the Green's function in the gauge--Goldstone sector}
\label{sec:solvingA4G}

\subsection{Gauge choice}

By inspection of the operator $\mathcal{M}^{-1}_{(A_\mu,G)}$ in~\eqref{eq:koperators}, we see that the gauge $\xi=\zeta=1$ is suitable in order to simplify the problem. In this case, we have
\begin{align}
\label{eq:MAmuGsimplify}
\mathcal{M}^{-1}_{(A_\mu,G);{\bf k}}(\varphi^{(1)})&=\begin{pmatrix}
M_{{\bf k}}^{-1}(\varphi^{(1)})\delta_{ij} & 0 & 0 \\
0 & M_{{\bf k}}^{-1}(\varphi^{(1)}) & \ 2g(\partial_z\varphi^{(1)}) \\
0 & 2g(\partial_z\varphi^{(1)}) & \ N_{{\bf k}}^{-1}(\varphi^{(1)}) \\
\end{pmatrix},
\end{align}
where now
\begin{align}
N_{{\bf k}}^{-1}(\varphi^{(1)})&=-\partial_z^2+{\bf k}^2+\alpha+\lambda\,{\varphi^{(1)2}}+\frac{3}{4}\lambda_6\,{\varphi^{(1)4}}+{g^2\,{\varphi^{(1)2}}}.
\end{align}
This gauge is particularly useful given that $M^{-1}_{{\bf k}}$ decouples from $\mathcal{M}^{-1}_{(A_\mu,G);\mathbf{k}}$ and has the same form as $\mathcal{M}^{-1}_{(\bar{\eta},\eta);{\bf k}}$, leading to a partial cancellation between the contributions of $M^{-1}_{\mathbf{k}}$ and $\mathcal{M}^{-1}_{(\bar{\eta},\eta);{\bf k}}$ to the effective action, as can be seen from Eq.~\eqref{eq:finaleffectiveaction}. Because of this cancellation, the effective action only contains three independent functional determinant terms corresponding to the scalar, the mixing $A_4$--Goldstone and the ghost sectors:
\begin{align}
\label{eq:sfinaleffectiveaction}
\Gamma^{(1)}[\varphi^{(1)}]&=S[\varphi^{(1)}]+\frac{\hbar}{2}\,V\int\frac{\D^3{\bf k}}{(2\pi)^3}\,\log\frac{\det\mathcal{M}^{-1}_{\hat{\Phi};{\bf k}}(\varphi^{(1)})}{\det\mathcal{M}^{-1}_{\hat{\Phi};{\bf k}}(0)}+\frac{\hbar}{2}\,V\int\frac{\D^3{\bf k}}{(2\pi)^3}\log\frac{\det\mathcal{M}^{-1}_{(A_4,G);{\bf k}}(\varphi^{(1)})}{\det\mathcal{M}^{-1}_{(A_4,G);{\bf k}}(0)}\notag\\
&+\frac{\hbar}{2}\,V\int\frac{\D^3{\bf k}}{(2\pi)^3}\log\frac{\det\mathcal{M}^{-1}_{(\bar{\eta},\eta);{\bf k}}(\varphi^{(1)})}{\det\mathcal{M}^{-1}_{(\bar{\eta},\eta);{\bf k}}(0)},
\end{align}
where $\mathcal{M}^{-1}_{(A_4,G);{\bf k}}$ can be read from Eq.~\eqref{eq:MAmuGsimplify}. Note that the sign in front of the ghost contribution is flipped, and the coefficient halved, due to the partial cancellation between gauge and ghost degrees of freedom mentioned above. From now on, we use $X\in\{\hat\Phi,(A_4,G),(\bar\eta,\eta)\}$.

Expanding $\varphi^{(1)}=\varphi_0+\hbar\delta\varphi$ as in the previous section, we have
\begin{align}
\label{eq:effectiveAction2}
\Gamma[\varphi^{(1)}]\ =\ &B + \frac{\hbar}{2}B^{(1)} + \hbar B^{(1)}_{\hat{\Phi};{\rm dis}}+\hbar\, B_{\hat{\Phi}}^{(1)}+\hbar\, B_{(A_4,	G)}^{(1)}-\frac{\hbar}{2}\, B_{(\bar{\eta},\eta )}^{(1)}+\frac{1}{2}\left(\hbar^2\,B_{\hat{\Phi}}^{(2)}+\hbar^2\,
B_{(A_4, G)}^{(2)}-\frac{\hbar^2}{2}\,B_{(\bar{\eta},\eta)}^{(2)}\right),
\end{align}
where
\begin{subequations}
	\begin{align}
	\label{eq:B1A4G}&B_{(A_4,G)}^{(1)}=\frac{1}{2}\,V\int\frac{\D^3{\bf k}}{(2\pi)^3}\log\frac{\det\mathcal{M}^{-1}_{(A_4,G);{\bf k}}(\varphi_0)}{\det\mathcal{M}^{-1}_{(A_4,G);{\bf k}}(0)}\;,\\
	&B_{(A_4,G)}^{(2)}=\frac{1}{2}\,V\int\frac{\D^3{\bf k}}{(2\pi)^3}\int\D z\,\delta\varphi(z)\,\frac{\delta}{\delta\varphi(z)}\left.\log\frac{\det\mathcal{M}^{-1}_{(A_4,G);{\bf k}}(\varphi)}{\det\mathcal{M}^{-1}_{(A_4,G);{\bf k}}(0)}\right|_{\varphi_0},
	\end{align}
\end{subequations}
and all remaining terms are unchanged with respect to Section~\ref{sec:Setup}. Correspondingly, the equation of motion for the corrected bounce takes the form Eq.~\eqref{eq:EoMbounce1} with the explicit potential term 
\begin{align}
U'_{\rm eff}(\varphi^{(1)};z)&=U'(\varphi^{(1)};z)+\hbar\,\Pi_
{\hat{\Phi}}(\varphi_0;z)\,\varphi_0(z)+\hbar\,\Pi_
{(A_4,G)}(\varphi_0;z)\,\varphi_0(z)-\frac{\hbar}{2}\,\Pi_
{(\bar{\eta},\eta)}(\varphi_0;z)\,\varphi_0(z),
\end{align}
where $\Pi_
{\hat{\Phi}}(\varphi_0;z)\,\varphi_0(z),\;\Pi_
{(\bar{\eta},\eta)}(\varphi_0;z)$ are given in Eqs.~\eqref{eq:Piphiall}, while
\begin{align}
&\Pi_
{(A_4,G)}(\varphi_0;z)\,\varphi_0(z)={\frac{1}{2}}\int\frac{\D^3{\bf k}}{(2\pi)^3}\,\tr \left[\mathcal{M}_{(A_4,G);{\bf k}}(\varphi_0;z)\left.\left(\frac{\partial
	\mathcal{M}^{-1}_{(A_4,G);{\bf k}}(\varphi^{(1)};z)}{\partial\varphi^{(1)}(z)}\right)\right|_{\varphi_0}\right].
\end{align}
It can be checked that Eq.~\eqref{eq:B2Piphi} is also valid for the $(A_4,G)$ sector. With the present gauge choice, we readily write
\begin{equation}
\left.\frac{\partial}{\partial \varphi}\mathcal{M}^{-1}_{(A_4,G);{\bf k}}(\varphi;z)\right|_{\varphi_0} = \begin{pmatrix}
2g^2\varphi_0(z) & -2g\partial_z \\
-2g\partial_z & 2\lambda\varphi_0(z) + 3\lambda_6\varphi_0^3(z) + 2g^2\varphi_0(z)
\end{pmatrix}.
\end{equation} 
Substituting the above expression into Eq.~\eqref{eq:PiA4G}, we have 
\begin{align}
\Pi_{(A_4, G)}(\varphi_0;z)\varphi_0(z) =& \frac{1}{2}\int \frac{\d^3 \bf{k}}{(2\pi)^3} \bigg( 2g^2\varphi_0(z)\mathcal{M}_{(A_4,A_4)}(z) - 4g\partial_z \mathcal{M}_{(A_4,G)}(z)\notag\\
& + (2g^2\varphi_0+2\lambda\varphi_0(z) + 3\lambda_6\varphi_0^3(z))\mathcal{M}_{(G,G)}(z)\bigg).
\label{eq:tapoleA4Gauge}
\end{align}

The correction to the bounce is now
\begin{align}\label{eq:deltavarphispecialgauge}\begin{aligned}
 \delta\varphi(z)=&\,\frac{1}{\hbar}{\cal M}_{\hat\Phi}(\varphi_0;z)(\Box\varphi_0-U'(\varphi_0;z))\\
 &-{\cal M}_{\hat\Phi}(\varphi_0;z)\left(\Pi_{\hat\Phi}(\varphi_0;z)\varphi_0(z)+\Pi_{(A_4, G)}(\varphi_0;z)\varphi_0(z)-\frac{1}{2}\Pi_{(\bar\eta,\eta)}(\varphi_0;z)\varphi_0(z)\right).
\end{aligned}\end{align}

The final expression for the tunnelling rate per unit volume then is 
\begin{align}
\label{eq:decayratespecialgauge}\begin{aligned}\frac{\gamma}{V}=&\,\left(\frac{B}{2\pi\hbar}\right)^2{\sqrt\frac{\alpha^5}{{}{|\lambda_0|}}}\,\exp\left[-\frac{1}{\hbar}\,\left({\cal B}^{(0)}+\hbar\,{\cal B}^{(1)}+\hbar^2\,{\cal B}^{(2)}\right)\right],\\
{\cal B}^{(0)}=&\,B,\\
{\cal B}^{(1)}=&\,\,B_{\hat{\Phi}}^{(1)}+
B_{(A_4,G)}^{(1)}-\frac{1}{2}\,B_{(\bar{\eta},
\eta)}^{(1)},\\
{\cal B}^{(2)}=&\,-{B^{(2)}},\\
\end{aligned}\end{align}
where 
\begin{align}\label{eq:B2specialgauge}
{B^{(2)}}=-\frac{1}{2}\,\left(\,B_{\hat{\Phi}}^{(2)}+\,B_{(A_4,G)}^{(2)} - \frac{1}{2}\,B_{(\bar{\eta},\eta)}^{(2)}\right)-\frac{1}{2\hbar}B^{(1)}.
\end{align}
	
Having collected these formulae, we see that, in the planar-wall limit and in comparison with the purely scalar case, the main complication in the calculation for the theory with additional gauge and Goldstone fields is due to the two-by-two matrix structure of the Green's function $\mathcal{M}_{(A_4,G);\mathbf{k}}(\varphi_0;z)$. Once more, what we need is to compute the coincident Green's functions and associated resolvents, which in turn determine the tadpole contributions through Eqs.~\eqref{eq:PiPhi}, \eqref{eq:PiEta} and \eqref{eq:tapoleA4Gauge}. The $B^{(1)}_X$  corrections follow from 
\eqref{eq:functionalDetDeformation} and the $B^{(2)}$ correction by means of Eqs.~\eqref{eq:B2specialgauge}, \eqref{eq:B10}, \eqref{eq:deltavarphispecialgauge} and \eqref{eq:B2Piphi}.
In the following subsections we focus on the methods used to calculate the coincident Green's functions.

\subsection{\label{subsec:hom}Solving for the Green's functions in the homogeneous background approximation}

For comparison with the results accounting for the full gradient effects and to facilitate the numerical implementation of the renormalization procedure detailed in Section \ref{sec:ren}, it is useful to collect approximate results for the coincident Green's functions ${\cal M}_{X;{\bf k}}(\varphi_0;z,z)$ obtained when the gradients of the background are neglected. We will refer to these approximations as ``homogeneous Green's functions'' and denote them by  ${\cal M}_{X;{\bf k};{\rm hom}}(\varphi_0;z,z)$. The idea is to solve for the Green's functions in a homogeneous background $\phi$, and at the end substitute $\phi$ by the bounce $\varphi_0$.
 
We start by representing the differential operators ${\cal M}^{-1}_{X;{\bf k}}(\phi;z)$ in terms of generalized matrices  ${\cal M}^{-1}_{X;{\bf k}}(\phi;z,z')=\delta(z'-z){\cal M}^{-1}_{X;{\bf k}}(\phi;z)$. With the chosen gauge-fixing and for the constant background $\phi$, the operators ${\cal M}^{-1}_{X;{\bf k}}(\phi;z,z')$  are diagonal with respect to the  discrete matrix structure in Eq.~\eqref{eq:koperators}, and one can write
\begin{align}
\label{eq:Mhom0}
  {\cal M}^{-1}_{X;{\bf k};{\rm hom}}(\phi;z,z')=\int\frac{\d k_4}{2\pi}\,  \e^{\i k_4(z-z')} ({\bf k}^2+k_4^2+{\bf m^2_X}(\phi)),
\end{align}
where ${\bf m^2_X}(\phi)$ is a diagonal matrix containing the background-dependent effective masses in the field sector labelled by $X$. The mass matrices following from \eqref{eq:koperators} in the $\xi=\zeta=1$ gauge are given next:
\begin{align}\label{eq:m2X}\begin{aligned}
 m^2_{\hat\Phi}(\phi)=&\,\alpha+3\lambda\phi^2+\frac{15\lambda_6}{4}\phi^4,\\
  {\bf m^2_{(A_4,G)}(\phi)}=&\,\left(\begin{array}{cc}
                          g^2\phi^2 & 0\\
                         0& \alpha+\lambda\phi^2+\frac{3\lambda_6}{4}\phi^4+g^2\phi^2
                         \end{array}\right),
\\
  m^2_{(\bar\eta,\eta)}(\phi)=&\,g^2\phi^2.
\end{aligned}\end{align}
Eq.~\eqref{eq:Mhom0} immediately implies
\begin{align}
\label{eq:Mhom}
 {\cal M}_{X;{\bf k};{\rm hom}}(\phi;z,z')=\int\frac{\d k_4}{2\pi}\,  \e^{\i k_4(z-z')} \frac{1}{{\bf k}^2+k_4^2+{\bf m^2_X}(\phi)}.
\end{align}
Evaluating the former in the planar bounce background, we get the homogeneous approximations to the coincident Green's functions,
\begin{align}
\label{eq:Mhom2}
 {\cal M}_{X;{\bf k};{\rm hom}}(\phi;z,z)=\frac{1}{2\sqrt{{\bf k}^2+{\bf m^2_X}(\varphi_0(z))}}.
\end{align}
The resolvents follow by substituting ${\bf k}^2\rightarrow{\bf k}^2+s$, which allows to estimate the tadpoles from Eqs.~\eqref{eq:PiPhi}, \eqref{eq:PiEta} and \eqref{eq:tapoleA4Gauge}. The $B^{(1)}$ contributions in this approximation are related to the usual Coleman-Weinberg potential. Indeed, starting with Eq.~\eqref{eq:functionalDetDeformation} and using Eq.~\eqref{eq:Mhom}, one obtains
\begin{align}\label{eq:dethom}\begin{aligned}
&\frac{1}{2}V\int\frac{\d^3{\bf k}}{(2\pi)^3} \log\frac{\det {\cal M}^{-1}_{X;{\bf k};{\rm hom}}(\varphi_0)}{\det {\cal M}^{-1}_{X;{\bf k};{\rm hom}}(\varphi_0)}\\
=&-\frac12\tr\int \d^4 x\int_0^\infty \d s \int\frac{\d^4k}{(2\pi)^4}\left(\frac{1}{k^2+s+{\bf m^2_X}(\varphi_0(z))}-\frac{1}{k^2+s+{\bf m^2_X}(0)}\right)\\
=&-\frac12\tr\int \d^4x\int_0^\infty \d s \frac{\partial}{\partial s}\int\frac{\d^4k}{(2\pi)^4}\log\left(\frac{k^2+s+{\bf m^2_X}(\varphi_0)}{k^2+s+{\bf m^2_X}(0)}\right)\equiv\int \d^4x\, U^{(1)}_{{\rm CW},X}(\varphi_0).
\end{aligned}\end{align}
In the last line we have identified the $U^{(1)}_{{\rm CW},X}$ as the one-loop contribution of the sector $X$ to the Coleman-Weinberg potential, (normalized to be zero at the origin, as commented after Eq.~\eqref{eq:greenMatrices}):
\begin{align}\label{eq:VCW}\begin{aligned}
 U^{(1)}_{{\rm CW}}(\varphi)=&\,U^{(1)}_{{\rm CW},\hat\Phi}(\varphi)+U^{(1)}_{{\rm CW},(A_4,G)}(\varphi)+U^{(1)}_{{\rm CW},(\bar\eta,\eta)}(\varphi),\\
 U^{(1)}_{{\rm CW},X}(\varphi)=&\,\frac{1}{2}\tr\int\frac{\d^4k}{(2\pi)^4}\log\left(\frac{k^2+{\bf m^2_X}(\varphi)}{k^2+{\bf m^2_X}(0)}\right).
\end{aligned}\end{align}
When using a cutoff regularization with cutoffs $\Lambda_s^2,\Lambda$ for the $s$ and $k$ integrals, the fact that in the final step in Eq.~\eqref{eq:dethom} one may ignore the contributions at the boundary $s=\infty$ can be justified by taking the limits of large $\Lambda_s,\Lambda$ while maintaining $\Lambda_s\gg\Lambda$. It can be checked that the tadpole contributions satisfy
 \begin{align}\label{eq:Pihom}
  \Pi_{X;{\rm hom}}(\varphi;z)\varphi(z)=\frac{\partial U^{(1)}_{{\rm CW},X}(\varphi(z))}{\partial\varphi(z)}.
 \end{align}


\subsection{\label{subsec:exactmethod}Solving the Green's functions \texorpdfstring{$\mathcal{M}_{(A_4,G);\mathbf{k}}(\varphi_0;z)$}{M(A4,G)(phi0)} directly for \texorpdfstring{$\mathbf{k}^2\lesssim g\partial_z\varphi_0$}{k^2< gphi0'}}

{For large momenta, the gradients of the bounce appearing in the off-diagonal components in Eq.~\eqref{eq:MAmuGsimplify}---which were neglected in the previous section---are sub-dominant and can be handled perturbatively. However} for values $k^2\lesssim g\partial_z\varphi_0$, the gradients become more relevant. Therefore, we take a specific approach to directly compute the Green's functions for low momenta, for which perturbative and iterative methods break down. 

In the following discussion, we denote $\mathcal{M}^{-1}_{(A_4,G);\mathbf{k}}(\varphi_0;z)$ as $\mathcal{M}^{-1}(z)$ for brevity. 
The system to be solved is
\begin{align}
\mathcal{M}^{-1}(z)\mathcal{M}(z,z')=\delta (z-z')\mathbbm{1}_2,
\label{eq:Green-A4G}
\end{align}
and we denote the elements of the solution that we seek after as
\begin{equation}
\label{eq:matrixM}
\mathcal{M}(z,z') = \begin{pmatrix}
M_{11}(z,z') &\quad  M_{12}(z,z')\\[6pt]
M_{21}(z,z') &\quad M_{22}(z,z')
\end{pmatrix}
.
\end{equation}
In order to obtain numerical solutions, we express each of the ${M}_{ij}(z,z')$ in terms of two functions ${M}^{L}_{ij}(z),{M}^R_{ij}(z)$ as
\begin{align}
{M}_{ij}(z,z') = \Theta(z-z'){M}^R_{ij}(z) + \Theta(z'-z){M}^L_{ij}(z),
\end{align}
where $\Theta$ is the Heavyside step-function.
 We can then solve Eq.~\eqref{eq:Green-A4G} as an ordinary differential equation for fixed values of $z^\prime$. For each component ${M}_{ij}$ we impose four boundary conditions, namely  ${M}_{ij}^L(-\infty)= 0$ and ${M}_{ij}^R(\infty)= 0$, together with the matching condition of continuity, ${M}_{ij}^L(z')={M}_{ij}^R(z')$, and of the jump in the derivative ${M'}_{ij}^{L}(z')-{M'}_{ij}^{R}(z')=1/(1-z'^2)$, which follows from integrating Eq.~\eqref{eq:Green-A4G} over $z$ around the singularity at $z=z'$.  It is useful to compactify the $z$-coordinate with the transformation 
 \begin{align}\label{eq:uz}
  u=\tanh(z),
 \end{align}
and in the rest of this section the dependence on the functions will be assumed to be on the variable $u$.

\begin{figure}[tbph]
	\centering
	\includegraphics[clip, scale=0.82]{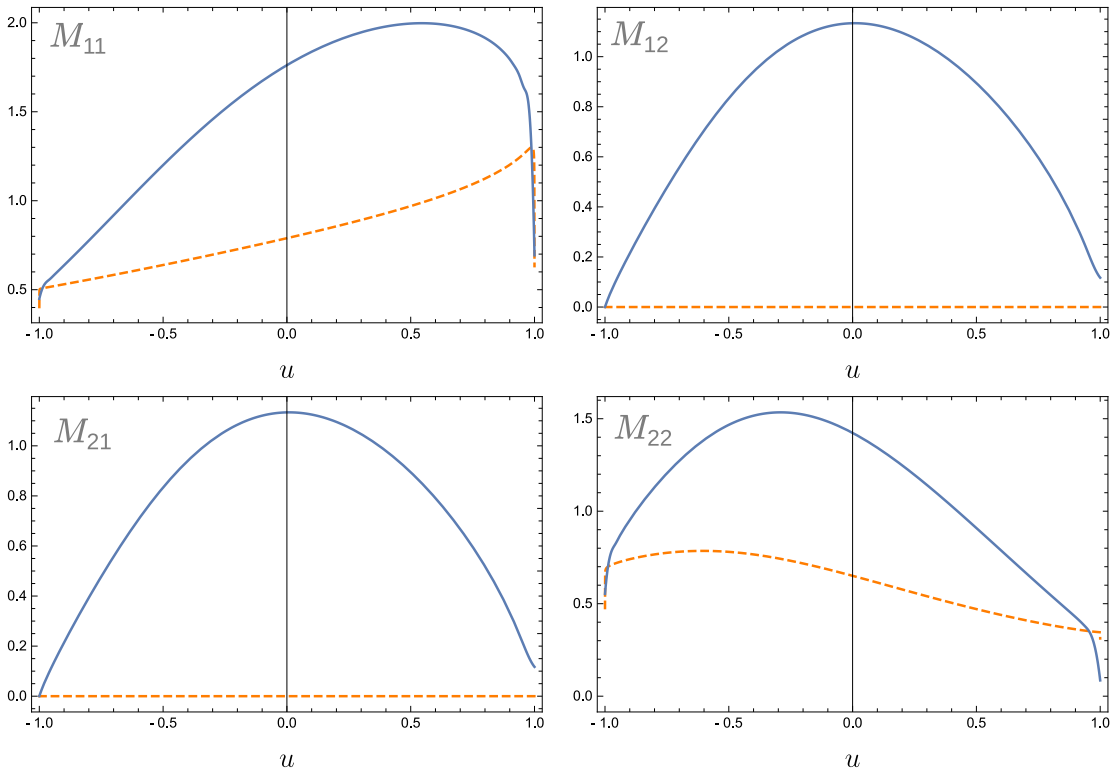}
	\captionsetup{justification=raggedright,singlelinecheck=false}
	\caption{For $k=0.3$ and with the couplings fixed as in Eq.~\ref{eq:benchmarkParams2}, comparison of the coincident limits of the direct solution to \eqref{eq:Green-A4G} (solid), and of the solution obtained when ignoring background gradients in ${\cal M}^{-1}_{(A_4,G)}(\varphi_0)$ (dashed). The graphs are labelled according to the matrix notation of Eq.~\eqref{eq:matrixM}.}
	\label{fig:coinGreenk03Axact}
\end{figure}

\begin{figure}[tbph]
	\centering
	\includegraphics[clip, scale=0.82]{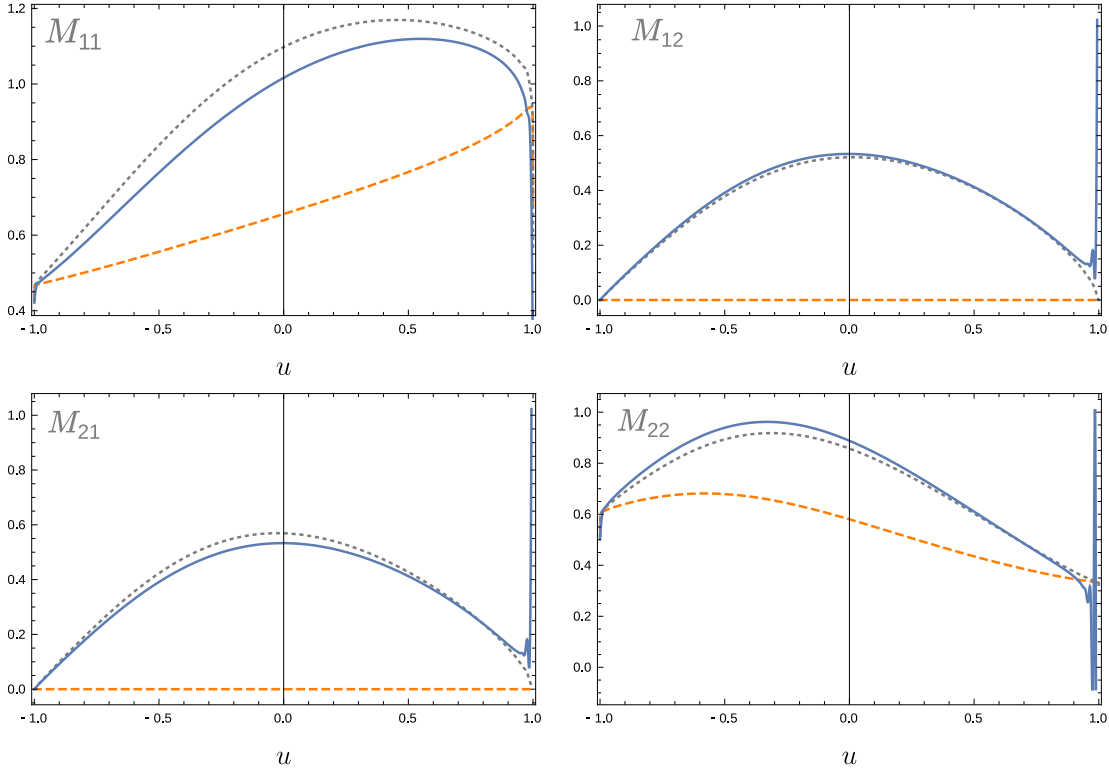}
	\captionsetup{justification=raggedright,singlelinecheck=false}
	\caption{For $k=0.5$ and with the couplings fixed as in Eq.~\ref{eq:benchmarkParams2}, comparison of the coincident limits of the solutions  to the system in \eqref{eq:Green-A4G} obtained when directly solving the full equation (solid), when ignoring background gradients in ${\cal M}^{-1}_{(A_4,G)}(\varphi_0)$ (dashed), and when using the perturbative numerical treatment of Section \ref{subsec:perturbativemethod} (dotted). The graphs are labelled with the matrix notation of Eq.~\eqref{eq:matrixM}.}
	\label{fig:coinGreenIterAllk05Axact}
\end{figure}

For the purpose of the present calculations, as explained in Section \ref{subsec:Green} we only need the coincident limit for the Green's functions. We have calculated the latter numerically, taking 1000 points between $(-1,1)$ for the compactified coordinate $u'=\tanh(z')$, and  solving for each $z'$ the matrix-valued differential equation~\eqref{eq:Green-A4G} as a function of $z$.  With this, one can evaluate the coincident Green's function for each matrix component and each of the chosen values of $z'$,  and the result can be interpolated in $z$. To deal with the effect of the $\delta$ function in Eq.~\eqref{eq:Green-A4G} we separate the equation for each matrix component into two differential equations, one for $z<z'$ involving the functions ${M}^L_{ij}(u)$, and one for $z>z'$ involving ${M}^R_{ij}(u)$. For each component, we have thus two differential equations, which can be solved numerically using the four boundary conditions detailed above. An example solution for a low value of $k=0.3$, using the parameter choices of Eq.~\eqref{eq:benchmarkParams2}, is displayed in Figure~\ref{fig:coinGreenk03Axact} (solid line) compared to the solution obtained when ignoring  gradients of the background in ${\cal M}^{-1}$ (dashed line). Note how the off-diagonal components are comparable to the diagonal ones, while neglecting background gradients in ${\cal M}^{-1}$ leads to vanishing  ${M}_{12}, {M}_{21}$. Figure~\ref{fig:coinGreenIterAllk05Axact} shows results for the value of $k=0.5$ where, with the chosen numerical implementation, the exact solution begins to have problems fulfilling the boundary conditions at the right edge of the domain. Nevertheless, for such large values of $k$ (leading to small ${M}_{12}, {M}_{21}$ in relation to ${M}_{11}, {M}_{11}$) one can start to use a perturbative treatment detailed in the following section, which leads to the dotted curves.  Although $k=0.5$ is at the margin of the validity of either method, in general the direct and perturbative estimates agree better with each other than with the approximation obtained by neglecting gradients.

\subsection{\label{subsec:perturbativemethod}Solving the Green's functions $\mathcal{M}_{(A_4,G);\mathbf{k}}(\varphi_0;z)$  iteratively for $\mathbf{k^2}\gtrsim g\partial_z\varphi_0$}

In this regime, the off-diagonal elements of $\mathcal{M}^{-1}(z)$ are small compared
to the diagonal ones.
Therefore, we decompose $\mathcal{M}^{-1}(z)$ as $\mathcal{M}^{-1}(z)=\mathcal{M}^{-1}_0(z)+\delta\mathcal{M}^{-1}(z)$ with
\begin{align}
\mathcal{M}^{-1}_0(z)=\begin{pmatrix}
M_{{\bf k}}^{-1}(\varphi_0(z)) & 0 \\
0 & N^{-1}_{{\bf k}}(\varphi_0(z))
\end{pmatrix}
\end{align}
and 
\begin{align}
\delta\mathcal{M}^{-1}(z)=\begin{pmatrix}
0 & 2g(\partial_z\varphi_0)\\
2g(\partial_z\varphi_0) & 0
\end{pmatrix}
\end{align}
as a perturbation. To set up an iterative solution, we let $\epsilon\sim\delta\mathcal{M}$ be a bookkeeping device for tracking the order of the expansion $\mathcal{M}=\mathcal{M}^{(0)} + \epsilon \mathcal{M}^{(1)} + \epsilon^2 \mathcal{M}^{(2)} + \cdots$. Then Eq. \eqref{eq:Green-A4G} can be written as
\begin{align}
(\mathcal{M}_0^{-1}(z)+\delta\mathcal{M}^{-1}(z))(\mathcal{M}^{(0)} + \epsilon \mathcal{M}^{(1)} + \epsilon^2 \mathcal{M}^{(2)} + \cdots )=\delta(z-z')\mathbbm{1},
\end{align}
which leads to
\begin{align}\label{eq:solNiterA4G}
\begin{aligned}
\mathcal{M}^{-1}_0(z)\mathcal{M}^{(0)}(z,z') &= \delta(z-z'),\\
\mathcal{M}^{-1}_0(z)\mathcal{M}^{(1)}(z,z')+\delta\mathcal{M}^{-1}(z)\mathcal{M}^{(0)}(z,z') &= 0,\\
\mathcal{M}^{-1}_0(z)\mathcal{M}^{(2)}(z,z')+\delta\mathcal{M}^{-1}(z)\mathcal{M}^{(1)} (z,z')&= 0,\\
&\vdots\\
\mathcal{M}^{-1}_0(z)\mathcal{M}^{({n+1})}(z,z')+\delta\mathcal{M}^{-1}(z)\mathcal{M}^{(n)}(z,z') &= 0. 
\end{aligned}\end{align}

\begin{figure}[htbp]
	\includegraphics[scale=.74,clip, trim= 0cm 0 0 0]{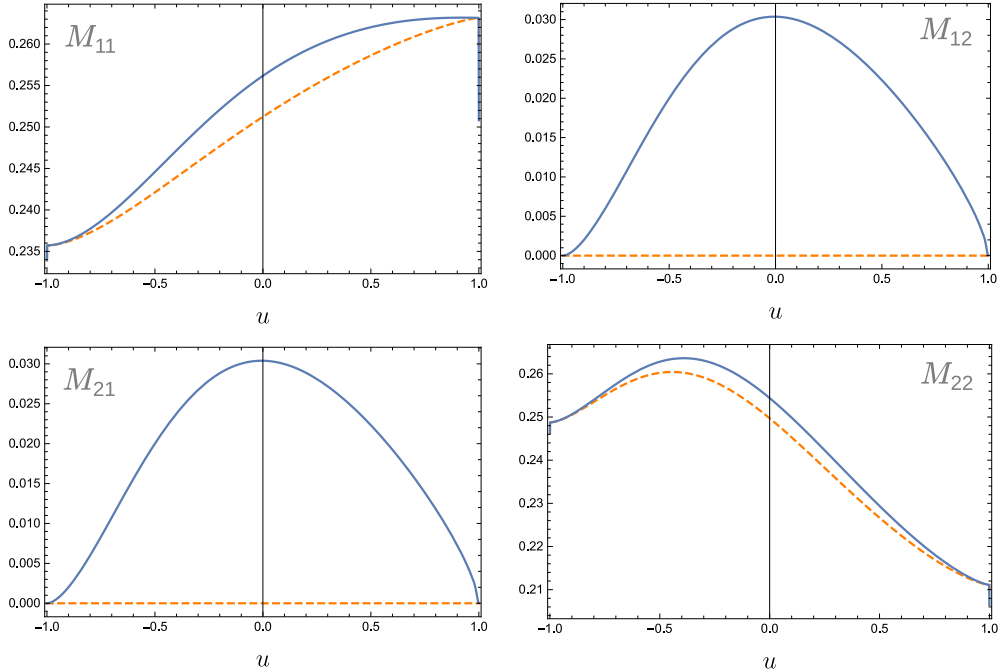}
	\captionsetup{justification=raggedright,singlelinecheck=false}
	\caption{Numerical results for $\mathcal{M}_{(A_4,G);{\bf k}}$ in the gauge $\xi=1$, $\zeta=1$ and $k=1.9$ for the benchmark parameters in Eqs.~\eqref{eq:benchmarkParams2}, at coincident limit in the background of the numerical bounce $\varphi_0$. The solid lines are obtained through the iterative strategy, while the dashed lines are the results at zeroth order ($\sim\mathcal{O}(\epsilon^0)$), in which background gradients in ${\cal M}_{A_4 G}^{-1}(\varphi_0)$ are ignored.}
	\label{fig:lowerBlockbounce}
\end{figure}

In the numerical implementation, we stop the iterative method when the difference between the results at order $\mathcal{O}(\epsilon^n)$ and $\mathcal{O}(\epsilon^{n+1})$ becomes less than ($10^{-5}$). The general behaviour is that the results for smaller $k$ converge more slowly and one needs more iterations, while for higher $k$, the solutions converge faster and moreover approach a solution that does not deviate too much from the solution where gradients are neglected. This is expected given that the gradients become less relevant for large momenta. Another feature of the perturbative expansion is that, although the Green's function must be symmetric, i.e. $M_{12}=M_{21}$, this property is not strictly maintained at every step in the perturbative expansion. This can then be used as a check of the reliability of the perturbative results. For example,  the result for $k=0.5$ shown in the dotted graphs in  Figure~\ref{fig:coinGreenIterAllk05Axact} shows a slight deviation from this symmetry, indicating as advertised earlier that the perturbative expansion starts to fail. Figure~\ref{fig:lowerBlockbounce} shows results for the Green's function in the background of the bounce $\varphi_0$---computed numerically---for $k=1.9$ and for the choices of parameters of Eq.~\eqref{eq:benchmarkParams2}. The perturbative expansion has converged after 14 iterations, and one recovers $M_{21}=M_{21}$ to much better accuracy than in Figure~\eqref{fig:coinGreenIterAllk05Axact}. The results at zeroth order ($\sim\mathcal{O}(\epsilon^0)$, dashed lines) are shown for comparison as well.

\section{Renormalization} 
\label{sec:ren}

The coincident Green's functions correspond to expectation values of composite operators, which are divergent and require renormalization. The divergences appear when computing ${\cal M}_{X}(\varphi_0;x',x)$ from Eq.~\eqref{eq:Mk}: While the ${\cal M}_{X;{\bf k}}(\varphi_0;z,z)$ computed as in the previous sections are finite, integral over $\bf k$ is not convergent. We will use a cutoff regulator, and remove the divergent contributions by means of counterterms.  Even though our calculations of the effective action include some two-loop contributions, it turns out that it suffices to compute the one-loop counterterms. As usual, the counterterms for the effective action are just local polynomials of the fields and their derivatives. Following methods applied in previous work~\cite{Ai:2018guc}, the counterterms that do not involve field derivatives---which will be referred to as ``coupling counterterms''---can be calculated simply by evaluating the effective action at a homogeneous field configuration, rather than at the bounce. As it was shown in section \ref{subsec:hom}, the one-loop result is related to the Coleman--Weinberg effective potential, which can be calculated analytically. For each renormalizable interaction, we need one counterterm, while we also anticipate, given the non-renormalizable nature of the $\phi^6$-theory, that a dimension-eight counterterm will be necessary.
 
For a scalar field with gauge interactions, it is known that one-loop corrections give rise to a logarithmic divergence in the scalar two-point function, which can be subtracted through a wave-function renormalization. The latter can be calculated analytically using a gradient expansion of the effective action. As said before, we do not rely on such an expansion to compute the regularized Green's functions prior to renormalization, so that our renormalized result for the effective action contains all the derivative corrections at the chosen truncation of the loop expansion. 

In summary, we consider a counterterm Lagrangian of the form
\begin{align}
\label{eq:Lct}
\mathcal{L}_{\rm ct}[\varphi] = \frac{1}{2}\delta Z (\partial\varphi)^2+\frac{\delta\alpha}{2}\varphi^2 + \frac{\delta\lambda}{4}\varphi^4 + \frac{\delta\lambda_6}{8}\varphi^6 + \frac{\delta \lambda_8}{16}\varphi^8.
\end{align}

The counterterms can be separated into a sum of terms, each renormalizing the contributions from the different sectors $X=\hat\Phi,(A_4,G),(\bar\eta,\eta)$ in the effective action:
\begin{align}\begin{aligned}
\mathcal{L}_{\rm ct}[\varphi] = \sum_X\mathcal{L}_{{\rm ct},X}[\varphi]=\sum_X \left(\frac{1}{2}\delta Z_X (\partial\varphi)^2+\frac{\delta\alpha_X}{2}\varphi^2 + \frac{\delta\lambda_X}{4}\varphi^4 + \frac{\delta\lambda_{6,X}}{8}\varphi^6 + \frac{\delta \lambda_{8,X}}{16}\varphi^8\right).
\end{aligned}\end{align}

We then define the renormalized one-loop contribution ${\cal B}^{(1){\rm ren}}$ contributing exponentially to the decay rate in Eq.~\eqref{eq:decayratespecialgauge} as:
\begin{align}\begin{aligned}
\label{Gamma:ren}
{\cal B}^{(1){\rm ren}}=&\,B^{(1){\rm ren}}_{\hat\Phi}+B^{(1){\rm ren}}_{(A_4,G)}-\frac{1}{2} B^{(1){\rm ren}}_{(\bar\eta,\eta)}\\
=&\, B^{(1)}_{\hat\Phi}+B^{(1)}_{(A_4,G)}-\frac{1}{2} B^{(1)}_{(\bar\eta,\eta)} + \int {\textrm{d}^4 x} \;\mathcal{L}_{\rm ct}[\varphi_0]\equiv \sum_X g_X  B^{(1){\rm ren}}_X,
\end{aligned}\end{align}
where we have introduced numerical factors
\begin{align}\label{eq:gXs}
 g_{\hat\Phi}=g_{(A_4,G)}=-2g_{(\bar\eta,\eta)}=1,
\end{align}
while
\begin{align}
 g_X  B^{(1){\rm ren}}_X =g_X  B^{(1)}_X+\int \d^4x\, {\cal L}_{{\rm ct},X}[\varphi_0].
\end{align}
In the numerical implementation, it is useful to express the above renormalized contributions in terms of convergent integrals. Following Eqs.~\eqref{eq:B1}, \eqref{eq:B1A4G}, \eqref{eq:functionalDetDeformation} and introducing cutoffs $\Lambda_s$, $\Lambda$ for the integrations in $s$ and $\bf k$, respectively, one has:
\begin{align}\begin{aligned}
 g_X B^{(1){\rm ren}}_{X}=&\,-\frac{1}{2}\tr\int \d^4 x \int_0^{\Lambda_s^2} \d s \int_{B_\Lambda}\frac{\D^3{\bf k}}{(2\pi)^3}
\left(\mathcal{M}_{X;\sqrt{{\bf k}^2+s}}(\varphi_0;z,z)-\mathcal{M}_{X;\sqrt{{\bf k}^2+s}}(0;z,z)\right)\\
&+\int \d^4x\, {\cal L}_{{\rm ct},X}[\varphi_0],
\end{aligned}\end{align}
where $B_\Lambda$ denotes a three-dimensional ball of radius $\Lambda$. One may note that  $\mathcal{M}_{X;\sqrt{{\bf k}^2+s}}(0;z,z)$ can be obtained from the analytic results for the homogeneous resolvents $\mathcal{M}_{X;\sqrt{{\bf k}^2+s};{\rm hom}}$ in Eq.~\eqref{eq:Mhom2}. Furthermore, the homogeneous resolvent gives a real result when evaluated at the false vacuum $\varphi=0$. To isolate the contributions from gradient effects, one can add and subtract the real part of the homogeneous resolvent evaluated at the bounce background, ${\rm Re}\mathcal{M}_{X;\sqrt{{\bf k}^2+s};{\rm hom}}(\varphi_0;z,z)$. Furthermore, we may add and subtract a term containing the contribution of the $X$ sector to the wave-function divergence. For this we construct a kernel $K_{s,X}({\bf k})$ satisfying:
\begin{align}\label{eq:Ksdef}
   \Gamma[\varphi_0]\supset-\int\d^4x\,\frac{1}{2}\delta Z_X(\partial_\mu\varphi_0)^2=-\int\d^4x\left(\frac{1}{2}\int_0^{\Lambda_s^2} \d s \int_{B_\Lambda}\frac{\D^3{\bf k}}{(2\pi)^3} K_{s,X}({\bf k})(\partial_z\varphi_0)^2+{\rm finite}\right).
\end{align}
We then obtain
\begin{align}\begin{aligned}
 &g_X B^{(1){\rm ren}}_{X}=\,\\
 &-\frac{1}{2}\int \d^4 x \int_0^{\Lambda_s^2} \d s \int_{B_\Lambda}\frac{\D^3{\bf k}}{(2\pi)^3}
\left(\tr(\mathcal{M}_{X;\sqrt{{\bf k}^2+s}}(\varphi_0;z,z)\!-\!{\rm Re}\mathcal{M}_{X;\sqrt{{\bf k}^2+s};{\rm hom}}(\varphi_0;z,z))\!-\!K_{s,X}({\bf k})(\partial_z\varphi_0)^2\right)\\
&-\frac{1}{2}\int \d^4 x \int_0^{\Lambda_s^2} \d s \int_{B_\Lambda}\frac{\D^3{\bf k}}{(2\pi)^3}
\left(\tr({\rm Re}\mathcal{M}_{X;\sqrt{{\bf k}^2+s};{\rm hom}}(\varphi_0;z,z)-\mathcal{M}_{X;\sqrt{{\bf k}^2+s};{\rm hom}}(0;z,z))\right)\\
&-\frac{1}{2}\int \d^4 x \int_0^{\Lambda_s^2} \d s \int_{B_\Lambda}\frac{\D^3{\bf k}}{(2\pi)^3}K_{s,X}({\bf k})(\partial_z\varphi_0)^2+\int \d^4x\, {\cal L}_{{\rm ct},X}[\varphi_0].
\end{aligned}\end{align}
As shown in Section \ref{subsec:hom}, the terms in the second line involving the homogeneous resolvents simply give the one-loop Coleman-Weinberg potential (see Eqs.~\eqref{eq:Mhom2}, \eqref{eq:dethom}). Supplemented with the coupling counterterms in ${\cal L}_{\rm ct}$, which by construction are engineered to remove the divergences in the Coleman-Weinberg potential, one simply gets a contribution involving the renormalized $U^{(1){\rm ren}}_{{\rm CW},X}$. Thus, one obtains
\begin{align}
\label{eq:B1ren}
 &g_X B^{(1){\rm ren}}_{X}=\,g_X B^{(1){\rm ren,hom}}_{X}+g_X B^{(1){\rm ren,grad}}_{X},
 \end{align}
 with
 \begin{align}\label{eq:B1ren2}\begin{aligned}
 g_X B_X^{(1){\rm ren,hom}}=&\,V\int \d z \,{\rm Re}\,U^{(1){\rm ren}}_{{\rm CW},X}(\varphi_0(z)),\\
 g_X B_X^{(1){\rm ren,grad}}=&\,V\int \d z\,\left(-\frac{1}{2} \left(\int_0^{\Lambda_s^2} \d s \int_{B_\Lambda}\frac{\D^3{\bf k}}{(2\pi)^3}K_{s,X}({\bf k})\right)+\frac{1}{2}\delta Z_{X}\right)(\partial_z\varphi_0(z))^2\\
 &\hskip-2.5cm-\frac{1}{2}V\!\int\! \d z \!\int_0^{\Lambda_s^2}\! \!\d s \!\int_{B_\Lambda}\!\frac{\D^3{\bf k}}{(2\pi)^3}
\left(\tr(\mathcal{M}_{X;\sqrt{{\bf k}^2+s}}(\varphi_0;z,z)\!-\!{\rm Re}\mathcal{M}_{X;\sqrt{{\bf k}^2+s};{\rm hom}}(\varphi_0;z,z))\!-\!K_{s,X}({\bf k})(\partial_z\varphi_0(z))^2\right),
\end{aligned}\end{align}
where we have separated contributions captured by the homogeneous Green's functions from the gradient corrections. As $K_{s,X}(\bf k)$ generates the divergent contributions to the wave-function renormalization, all the above integrals remain finite by construction when taking the cutoffs to infinity, which is useful for the numerical calculations. As mentioned in section \ref{subsec:hom}. it is assumed that the cutoffs satisfy $\Lambda_s\gg\Lambda$.

Finally, it remains to renormalize the two-loop contributions $B^{(2)}$. As shown next, this can also be done in terms of the one-loop counterterms. First, recall that the contributions $B^{(2)}$ can be entirely calculated from the Green's functions ${\cal M}_{X;{\bf k}}$ and the tadpoles $\Pi_{X;{\bf k};z}(\varphi_0)\varphi_0(z)$, as follows from Eqs.~\eqref{eq:deltavarphispecialgauge} and \eqref{eq:B2Piphi}. Both $\delta\varphi$ and $B^{(2)}$ inherit divergences from the tadpoles $\Pi_{X}(\varphi_0;z)\varphi_0(z)$. As explained in Section \ref{subsec:Green}, the tadpoles correspond to functional derivatives of the one-loop contributions to the effective action. Then it follows that one can obtain renormalized tadpoles by adding derivatives of the counterterms in ${\cal L}_{\rm ct}$,
\begin{align}\label{eq:Piren}
( \Pi_{X}(\varphi_0;z)\varphi_0(z))^{\rm ren} &= \Pi_{X}(\varphi_0;z)\varphi_0(z) -\delta Z_X \Box \varphi_0(z)+{\delta\alpha_X}\,\varphi_0(z) + {\delta\lambda_X}\,\varphi_0(z)^3 \notag\\
&+ \frac{3\delta\lambda_{6,X}}{4}\,\varphi_0(z)^5 + \frac{\delta \lambda_{8,X}}{2}\,\varphi_0(z)^7.
\end{align}
Using the renormalized tadpoles $ \Pi_{X}(\varphi_0;z)\varphi_0(z))^{\rm ren} $ in Eq.~\eqref{eq:deltavarphispecialgauge} one obtains finite  $\delta\varphi_0$, and doing the same in Eq.~\eqref{eq:B10}, \eqref{eq:B2Piphi}  the resulting values of $B^{(1)}$, $B^{(2)}_X$ are also renormalized. Explicitly,
\begin{align}
\label{eq:deltaB2ren}
\begin{aligned}
 \delta\varphi^{\rm ren}(z)=&\,\frac{1}{\hbar}{\cal M}_{\hat\Phi}(\varphi_0;z)(\Box\varphi_0-U'(\varphi_0;z))-{\cal M}_{\hat\Phi}(\varphi_0;z)\sum_X g_X (\Pi_{X}(\varphi_0;z)\varphi_0(z))^{\rm ren},\\
 B^{(1){\rm ren}}=&V\int \d z \,\delta\varphi^{\rm ren}(z)(-\Box\varphi_0+U'(\varphi_0;z)),\\
  B^{(2){\rm ren}}_X=&\,V\int dz\,\delta\varphi^{\rm ren}(z)\,(\Pi_X(\varphi_0;z) \varphi_0(z))^{\rm ren},\\
\end{aligned}\end{align}
so that the total two-loop exponential contribution to the decay rate, ${\cal B}^{(2){\rm ren}}$,  (see Eqs.~\eqref{eq:decayratespecialgauge}, \eqref{eq:B2specialgauge}) can be written as
\begin{align}\label{eq:B2rentotal}
  {\cal B}^{(2){\rm ren}}=&- { B}^{(2){\rm ren}}=-\frac{1}{2}\left({B}^{(2){\rm ren}}_{\hat\Phi}+{B}^{(2){\rm ren}}_{(A_4,G)}-\frac{1}{2}{B}^{(2){\rm ren}}_{(\bar\eta,\eta)}\right)-\frac{1}{2\hbar}B^{(1){\rm ren}}.
\end{align}

In the next subsections we obtain the coupling and wave-function counterterms separately.

\subsection{Coupling counterterms}

The coupling counterterms can be extracted from the divergent contributions to the Coleman--Weinberg potential, which is related to the effective action evaluated at a constant value of the field.
The starting point is Eq.~\eqref{eq:sfinaleffectiveaction} evaluated at a homogeneous field configuration $\phi$,
\begin{align}
\label{CW-action}
\begin{aligned}
\Gamma_{\rm CW}[\phi] = &\,\int\D^4 x\, U_{\rm CW}=S[\phi]+\frac{\hbar}{2}\,V\int\frac{\D^3{\bf k}}{(2\pi)^3}\,\log\frac{\det\mathcal{M}^{-1}_{\hat{\Phi};{\bf k}}(\phi)}{\det\mathcal{M}^{-1}_{\hat{\Phi};{\bf k}}(0)}\\
&+\frac{\hbar}{2}\,V\int\frac{\D^3{\bf k}}{(2\pi)^3}\log\frac{\det\mathcal{M}^{-1}_{(A_4,G);{\bf k}}(\phi)}{\det\mathcal{M}^{-1}_{(A_4,G);{\bf k}}(0)}+\frac{\hbar}{2}\,V\int\frac{\D^3{\bf k}}{(2\pi)^3}\log\frac{\det\mathcal{M}^{-1}_{(\bar{\eta},\eta);{\bf k}}(\phi)}{\det\mathcal{M}^{-1}_{(\bar{\eta},\eta);{\bf k}}(0)}.
\end{aligned}\end{align}
Note that this evaluation of  Eq.~\eqref{eq:sfinaleffectiveaction} does not correspond to the value of effective action for homogeneous fields which would require a Maxwell
construction when $\phi$ is not equal to the global minimum.
In the present case, the logarithms of the determinants of the  Green functions
can be obtained as in Section ~\ref{subsec:hom}, giving the one-loop contributions to the Coleman--Weinberg potential as in Eqs.~\eqref{eq:dethom}, \eqref{eq:VCW}. We indicate this interpretation by the subscript CW. To regularize the divergences we perform the full integration over $k_4$ in Eq.~\eqref{eq:VCW},  while we restrict the integration over the momenta parallel to the bubble wall to a large three-dimensional ball $B_\Lambda$ of radius $\Lambda$.  Adding the non-derivative counterterms to Eq.~\eqref{eq:Lct}, the renormalized effective potential can be written as
\begin{align}
\label{U:ren:1}
U^{\rm ren}_{\textrm{CW}}(\phi) &=  U(\phi) +  U^{(1)}_{{\rm CW},\hat\Phi}(\phi) +  U^{(1)}_{{\rm CW},(A_4,G)} +  U^{(1)}_{{\rm CW},(\bar\eta,\eta)}+ \frac{\delta\alpha}{2}\phi^2 + \frac{\delta\lambda}{4}\phi^4 + \frac{\delta\lambda_6}{8}\phi^6 + \frac{\delta \lambda_8}{16}\phi^8,
\end{align}
where $U(\phi)$ is the tree-level potential, as in Eq.~\eqref{eq:Vphi1}. Separating the contributions to $ U^{(1)}_{{\rm CW},X}$ involving different components of the mass matrices $\bf m^2_X$ of Eq.~\eqref{eq:m2X}, one can write 
\begin{align}\label{U:ren:2}\begin{aligned}
U^{(1)}_{{\rm CW},\hat\Phi}=&\,I_1, &
U^{(1)}_{{\rm CW},(A_4,G)}=&\,\frac{1}{2}I_2+I_3, &
U^{(1)}_{{\rm CW},(\bar\eta,\eta)}=&\,\frac{1}{2}I_2,
\end{aligned}\end{align}
with
\begin{align}\label{U:ren:3}\begin{aligned}
I_1 &\equiv \frac{\hbar}{2}\int_{B_\Lambda} \frac{\D^3{\bf k}}{(2\pi)^3}\int_{-\infty}^\infty\frac{\D k_4}{2\pi}\log\frac{k_4^2+{\bf k}^2 + U''(\phi)}{k_4^2+{\bf k}^2+U''(0)},\\
I_2 &\equiv \hbar\int_{B_\Lambda} \frac{\D^3{\bf k}}{(2\pi)^3}\int_{-\infty}^\infty\frac{\D k_4}{2\pi}\log\frac{k_4^2+{\bf k}^2 + g^2\phi^2}{k_4^2+{\bf k}^2},\\
I_3 &\equiv \frac{\hbar}{2}\int_{B_\Lambda} \frac{\D^3{\bf k}}{(2\pi)^3}\int_{-\infty}^\infty\frac{\D k_4}{2\pi}\log\frac{k_4^2+{\bf k}^2 + \alpha + \lambda\phi^2 + \frac{3\lambda_6}{4}\phi^4+g^2\phi^2}{k_4^2+{\bf k}^2+\alpha}.
\end{aligned}\end{align}
The integration gives
\begin{align}
\label{eq:cwIntegrals}
\begin{aligned}
I_1 & = \hbar\left[\frac{\Lambda^2}{16\pi^2}U''(\phi) + \frac{1}{64\pi^2}(U''(\phi))^2\left(\frac{1}{2}+ \ln\frac{U''(\phi)}{4\Lambda^2}\right) \right] - (\phi\leftrightarrow 0)+{\cal O}\left(\frac{1}{\Lambda}\right),\\
I_2 &=2\hbar\left[\frac{\Lambda^2}{16\pi^2}g^2\phi^2 + \frac{1}{64\pi^2}g^4\phi^4\left(\frac{1}{2}+ \ln\frac{g^2\phi^2}{4\Lambda^2}\right) \right]+{\cal O}\left(\frac{1}{\Lambda}\right),\\
I_3 &= \hbar\left[\frac{\Lambda^2}{16\pi^2}\left(\alpha + \lambda\phi^2 + \frac{3\lambda_6}{4}\phi^4+g^2\phi^2\right) + \frac{1}{64\pi^2}\left(\alpha+\lambda\phi^2+\frac{3\lambda_6}{4}\phi^4+g^2\phi^2\right)^2\times\right.\\
&\qquad\left. \times\left(\frac{1}{2}+ \ln\frac{\alpha + \lambda\phi^2 + \frac{3\lambda_6}{4}\phi^4+g^2\phi^2}{4\Lambda^2}\right)\right] - (\phi\leftrightarrow 0)+{\cal O}\left(\frac{1}{\Lambda}\right).  
\end{aligned}
\end{align}
We identify the counterterms from the divergent pieces proportional to $\Lambda^2$ and $\log\Lambda$, which need to be subtracted. We use a ``minimal subtraction'' scheme in which no finite part is included in the counterterms, except for the introduction of a subtraction scale $\mu$ needed on dimensional grounds for the contributions involving $\log\Lambda$. We can thus build a finite, renormalized effective potential as
\begin{equation}
\label{U:ren:4}
U^{\rm ren}_{\rm CW} = U_{\rm CW}-U_{\rm ct} = U_{{\rm CW}}- \Lambda^2 C_1(\phi) - \log\left(\frac{\Lambda}{\mu}\right)C_2(\phi),
\end{equation}
where $U_{\rm CW}= U(\phi)+ I_1+I_2+I_3$, while $C_1$ and $C_2$ are the corresponding coefficients of the contributions proportional to $\Lambda^2,\log\Lambda$ obtained from $I_1,I_2$ and $I_3$.
Comparing Eq.~\eqref{U:ren:1} with~\eqref{U:ren:4}, we find the renormalization
constants
\begin{align}
\label{eq:counterTermsMS}
\begin{aligned}
\delta\alpha &= \frac{1}{8 \pi ^2}\left[\alpha\left(g^2+4\lambda \right) \log\left(\frac{\Lambda}{\mu}\right)-\Lambda ^2 \left(3 g^2+ 4\lambda \right)\right],\\
\delta\lambda &= -\frac{1}{8\pi^2}\left[9 \lambda_6 \Lambda^2 - \log\left(\frac{\Lambda}{\mu}\right) \left(9 \alpha\lambda_6 + 3g^4 + 2g^2\lambda + 10\lambda^2\right)\right],\\
\delta\lambda_6 &= \frac{3}{8\pi^2}\left(g^2 + 16\lambda\right)\lambda_6 \log\left(\frac{\Lambda}{\mu}\right),\quad
\delta\lambda_8 = \frac{117\lambda_6^2}{16\pi^2}\log\left(\frac{\Lambda}{\mu}\right).
\end{aligned}
\end{align}
Alternatively, one could impose particular renormalization conditions
on the effective potential as e.g. in Refs.~\cite{Garbrecht:2015oea,Garbrecht:2015yza,Ai:2018guc}. In the present setup, this would however lead to comparably complicated expressions for the counterterms, which is why we proceed with the minimal subtraction of Eq.~\eqref{eq:counterTermsMS}.

As discussed above, in order to define renormalized tadpoles $(\Pi_{X}(\varphi_0;z)\varphi_0(z))^{\rm ren}$ we need to separate the above counterterms into contributions that subtract the divergences of the loop corrections associated with the sector $X$, i.e. the divergences in $U^{(1)}_{{\rm CW},X}$. Proceeding as it has been done before for the total one-loop potential  $U^{(1)}_{{\rm CW}}$ one finds:
\begin{align}
\label{eq:counterTermsMS2}
\begin{aligned}
\delta\alpha_{\hat\Phi} &= \frac{1}{8 \pi ^2}\left[3\alpha\lambda\log\left(\frac{\Lambda}{\mu}\right)-3\lambda\Lambda ^2\right], \\
\delta\alpha_{(A_4,G)} &= \frac{1}{8 \pi ^2}\left[\alpha\left(g^2+\lambda \right) \log\left(\frac{\Lambda}{\mu}\right)-\Lambda ^2 \left(2 g^2+ \lambda \right)\right],\\
\delta\alpha_{(\bar\eta,\eta)} &= \frac{-g^2\Lambda ^2 }{8 \pi ^2},\\
\end{aligned}\end{align}
\begin{align}
 \begin{aligned}
\delta\lambda_{\hat\Phi} &= \frac{3}{16\pi^2} \left[\left(5 \alpha  \lambda _6+6 \lambda ^2\right) \log \left(\frac{\Lambda }{\mu }\right)-5 \lambda _6 \Lambda ^2\right],\\
\delta\lambda_{(A_4,G)} &= 
\frac{1}{16\pi^2} \left[\left(3 \alpha  \lambda _6+4 g^2 \lambda +4 g^4+2 \lambda ^2\right) \log \left(\frac{\Lambda }{\mu }\right)-3 \lambda _6 \Lambda ^2\right], \\
\delta\lambda_{(\bar\eta,\eta)} &= \frac{g^4}{8\pi^2} \log \left(\frac{\Lambda }{\mu }\right),
\end{aligned}\end{align}
\begin{align}
\begin{aligned}
\delta\lambda_{6,\hat\Phi} &=\frac{ 45}{8\pi^2} \lambda  \lambda _6 \log \left(\frac{\Lambda }{\mu }\right),&
\delta\lambda_{6,(A_4,G)} &= \frac{3 }{8\pi^2}\lambda _6 \left(g^2+\lambda \right) \log \left(\frac{\Lambda }{\mu }\right),&
\delta\lambda_{6,(\bar\eta,\eta)} &=0 ,
\end{aligned}\end{align}
\begin{align}
\begin{aligned}
\delta\lambda_{8,\hat\Phi} &= \frac{225}{32\pi^2} \lambda _6^2 \log \left(\frac{\Lambda }{\mu }\right),&
\delta\lambda_{8,(A_4,G)} &= \frac{9}{32\pi^2} \lambda _6^2 \log \left(\frac{\Lambda }{\mu }\right),&
\delta\lambda_{8,(\bar\eta,\eta)} &=0.
\end{aligned}
\end{align}
The loop corrections are smallest when choosing $\mu^2$ to be of order
of the numerators in the logarithms of Eqs.~(\ref{eq:cwIntegrals}).
Note that, of course, the couplings depend on the renormalization scale
$\mu$ as well.

\subsection{Wave-function renormalization}

For the wave-function renormalization, we follow a procedure analogous to the previous section. The aim is to obtain an analytic expression for the derivative corrections to the effective action containing divergent terms, and to define the wave-function counterterm through minimal subtraction.

As explained in Ref.~\cite{Ai:2018guc}, interactions with scalar fields do not lead to a cutoff-dependent wave-function renormalization at one-loop order. We therefore focus here on the corrections that arise from the interaction of the
Higgs field with the gauge-Goldstone sector. For the present purpose, we consider the effective action evaluated at a general inhomogeneous background field $\varphi$ with the same spherical symmetry as the bounce $\varphi_0$. In the following, we derive the wave-function renormalization using a covariant gradient expansion as in Refs.~\cite{Chan:1985ny,Cheyette:1985ue,Gaillard:1985uh,Chan1986,Cheyette:1987qz,Henning2016}.
We first write the logarithm of the determinant of the gauge-Goldstone operator by tracing over a basis of  plane waves, so that matrix elements are written in position space as
\begin{equation}
\label{logdet:insertion:momentumstates}
\Gamma\supset\frac12\,{\rm Tr}\log\mathcal{M}^{-1}_{(A_{\mu},G)}(\varphi) = \frac12\int {\rm d}^4x\frac{{\rm d}^4p}{(2\pi)^4}\, \e^{\i px}{\tr\,\log} \mathcal{M}^{-1}_{(A_{{\mu}},G)}(x)\e^{{-}\i px},
\end{equation}
where the trace ``$\tr$'' that remains is over the matrix structure in Eq.~\eqref{eq:MAmuGsimplify} {and where we have written the explicit argument $x$ for $\mathcal{M}^{-1}_{A_\mu,G}$ instead of $\varphi(x)$, which will be clearer for the following manipulations.} Note that here $\log \mathcal{M}^{-1}_{(A_{\mu},G)}(x)$ corresponds to a representation of the operator as a differential operator acting on functions, rather than as a matrix with two continuous indices, as was e.g. used in Eq.~\eqref{eq:logMk0}.
Considering that to first order one has
\begin{equation}
\e^{\i px}\partial_\mu \e^{{-}\i px} = \partial_\mu {-} \i p_\mu,
\end{equation}
then acting with the exponentials on the logarithm gives~\cite{Chan1986}
\begin{align}
\label{eq:V15}
\Gamma &\supset \frac{1}{2}\int {\rm d}^4 x \frac{{\rm d}^4 p}{(2\pi)^4}\tr\,\log\mathcal{M}^{-1}_{(A_\mu,G,p)}(x),\\
\mathcal{M}^{-1}_{(A_\mu,G,p)}(x) &= \begin{pmatrix}
((\Pi - p)^2 + m_A(\varphi))\delta_{\mu\nu} & 2g(\partial_\mu\varphi) \\
2g(\partial_\nu\varphi) & (\Pi - p)^2 + m^2_G(\varphi)
\end{pmatrix}
,
\end{align}
where $\Pi_\mu=-\i\partial_\mu$.
{We now insert an identity operator $\exp(-\Pi_\mu\frac{\partial}{\partial p_\mu})\exp(\Pi_\mu\frac{\partial}{\partial p_\mu})$ into Eq.~\eqref{eq:V15} and obtain
\begin{align}\label{eq:GammaPi}\begin{aligned}
\Gamma &\supset \frac{1}{2}\int {\rm d}^4 x\frac{{\rm d}^4 p}{(2\pi)^4}\tr e^{-\Pi\cdot\frac{\partial}{\partial p}} e^{\Pi\cdot\frac{\partial}{\partial p}}\log(\mathcal{M}^{-1}_{(A_\mu,G,p)}(x))\\
&=\frac{1}{2}\int {\rm d}^4 x\frac{{\rm d}^4 p}{(2\pi)^4}\tr e^{\Pi\cdot\frac{\partial}{\partial p}} \log(\mathcal{M}^{-1}_{(A_\mu,G,p)}(x))e^{-\Pi\cdot\frac{\partial}{\partial p}},
\end{aligned}\end{align}
where we have used the cyclic property of the trace (understood in the functional space).}
Using an expansion of the logarithm by adding and subtracting the identity operator, we can push the exponentials inside the logarithm and use the Baker-Campbell-Hausdorff formula to rewrite the operators as:
\begin{equation}
e^{\Pi\cdot\frac{\partial}{\partial p}}\mathcal{O}e^{-\Pi\cdot\frac{\partial}{\partial p}} = \exp\left({\rm ad}_{\Pi\cdot\frac{\partial}{\partial p}}\right)\mathcal{O} = \sum_{n=0}^\infty \frac{1}{n!}\left({\rm ad}_{\Pi\cdot\frac{\partial}{\partial p}}\right)^n\mathcal{O},
\end{equation}
where
\begin{equation}
\text{ad}_{\Pi\cdot\frac{\partial}{\partial p}}\mathcal{O} = \left[\Pi\cdot\frac{\partial}{\partial p},\mathcal{O}\right].
\end{equation}
Observing that $[\Pi\cdot\frac{\partial}{\partial p},\Pi_\mu] = 0$ and $[\Pi\cdot\frac{\partial}{\partial p},p_\mu] = \Pi_\mu$, one has the following formulae:
\begin{align}\label{eq:Pip}\begin{aligned}
e^{\Pi\cdot\frac{\partial}{\partial p}}(\Pi_\mu - p_\mu)e^{-\Pi\cdot\frac{\partial}{\partial p}} = &\,e^{\Pi\cdot\frac{\partial}{\partial p}}\Pi_\mu e^{-\Pi\cdot\frac{\partial}{\partial p}}- e^{\Pi\cdot\frac{\partial}{\partial p}}p_\mu e^{-\Pi\cdot\frac{\partial}{\partial p}}\\
= \Pi_\mu - p_\mu - \sum_{n=1}^\infty\frac{1}
{n!}\left(\ad_{\Pi\cdot\frac{\partial}{\partial p}}\right)^{n-1}\Pi_\mu
= &\, \Pi_\mu - p_\mu - \Pi_\mu - \sum_{n=2}^\infty\frac{1}
{n!}\left(\ad_{\Pi\cdot\frac{\partial}{\partial p}}\right)^{n-1}\Pi_\mu=-p_\mu.
\end{aligned}\end{align}
For an arbitrary function of $x$, one observes
\begin{align}
\label{eq:V21}
e^{\Pi\cdot\frac{\partial}{\partial p}}f(x)e^{-\Pi\cdot\frac{\partial}{\partial p}} \equiv \tilde{f}(x) = &  \sum_{n=0}^\infty \frac{1}{n!}\prod_{i=1}^n \ad_{\Pi_{\mu_i}}f(x)\frac{\partial}{\partial p_{\mu_i}}
= f(x)+{\delta} f(x),
\end{align}
where 
\begin{align}
{\delta} f(x) &= \sum_{n=1}^\infty \frac{1}{n!}\prod_{i=1}^n \ad_{\Pi_{\mu_i}}f(x)\frac{\partial}{\partial p_{\mu_i}}
=\sum_{n=1}^\infty \frac{(-i)^n}{n!}(\partial_{\mu_1}\partial_{\mu_2}\cdots\partial_{\mu_n} f(x))\frac{\partial^n}{\partial p_{\mu_1}\partial p_{\mu_2}\cdots\partial p_{\mu_n}}.
\label{eq:momentumShift}
\end{align}
One can act on the operator $\mathcal{M}^{-1}_{(A_\mu,G,p)}$ in Eq.~\eqref{eq:GammaPi} with the exponentials by using equations \eqref{eq:Pip} and \eqref{eq:V21}. Decomposing the resulting matrix operator $\widetilde{\mathcal{M}}^{-1}_{(A_\mu,G,p)}$ into a free piece, a piece coming from the effective masses and one from the gradients of the background, the result is
\begin{equation}
\widetilde{\mathcal{M}}^{-1}_{(A_\mu,G,p)}(x)
=e^{\Pi\cdot\frac{\partial}{\partial p}} \mathcal{M}^{-1}_{(A_\mu,G,p)}(x) e^{-\Pi\cdot\frac{\partial}{\partial p}} 
=\widetilde{\mathcal{M}}^{-1}_{0(A_\mu,G,p)}(x) + \widetilde{\mathcal{M}}^{-1}_{1(A_\mu,G,p)}(x) + \widetilde{\mathcal{M}}^{-1}_{2(A_\mu,G,p)}(x),
\end{equation}
where
\begin{align}\begin{aligned}
\widetilde{\mathcal{M}}^{-1}_{0(A_\mu,G,p)}(x) &= \begin{pmatrix}
(p^2+m_A^2)\delta_{\mu\nu} & 0 \\
0 & p^2 + m_G^2
\end{pmatrix},\\
\widetilde{\mathcal{M}}^{-1}_{1(A_\mu,G,p)}(x) &= \begin{pmatrix}
{\delta} m_A^2\,\delta_{\mu\nu} & 0 \\
0 & {\delta} m_G^2
\end{pmatrix},\\
\widetilde{\mathcal{M}}^{-1}_{2(A_\mu,G,p)}(x) &= \begin{pmatrix}
0 & 2g\,{\widetilde{\partial_\mu\varphi}} \\
2g\, {\widetilde{\partial_\mu\varphi}} & 0
\end{pmatrix}.
\end{aligned}\end{align}

Expanding the logarithm of the shifted operator about the free contribution, one gets:
\begin{align}
\tr \log\widetilde{\mathcal{M}}^{-1}_{(A_\mu,G,p)}(x) &= \tr\log\widetilde{\mathcal{M}}^{-1}_{0(A_\mu,G,p)}\nonumber\\
&\quad- \tr\sum_{m=1}^{\infty} \frac{(-1)^m}{m}\left( \widetilde{\mathcal{M}}_{0(A_\mu,G,p)}(\widetilde{\mathcal{M}}^{-1}_{1(A_\mu,G,p)} + \widetilde{\mathcal{M}}^{-1}_{2(A_\mu,G,p)})\right)^m,
\label{eq:logExpansion}
\end{align}
with 
\begin{equation}
\widetilde{\mathcal{M}}_{0(A_\mu,G,p)} = \begin{pmatrix}
\frac{\delta_{\mu\nu}}{p^2+m_A^2} & 0\\
0 & \frac{1}{p^2+m_G^2}
\end{pmatrix} \equiv \begin{pmatrix}
\Delta_{A,\mu\nu} & 0 \\
0 & \Delta_G
\end{pmatrix}.
\end{equation}

Now we assemble contributions to the wave-function renormalization by collecting terms proportional to a kinetic term. Let us first deal with the terms coming from $\widetilde{\mathcal{M}}^{-1}_{1}$, that is the contributions corresponding to $n=2$ in the expansion in Eq.~\eqref{eq:momentumShift} and $m=1$ in the expansion of the logarithm in Eq.~\eqref{eq:logExpansion}:
\begin{equation}
\Gamma \supset -\frac{1}{4}\int{\rm d}^4x \frac{{\rm d}^4 p}{(2\pi)^4}\left(\frac{\partial^2 \Delta_{A,\mu\mu}}{\partial p_\rho \partial p_\sigma}\partial_\rho\partial_\sigma m_A^2 + \frac{\partial^2 \Delta_{G}}{\partial p_\rho\partial p_\sigma}\partial_\rho\partial_\sigma m_G^2 \right),
\end{equation}
where
\begin{align}
\frac{\partial^2 \Delta_{A,\mu\mu}}{\partial p_\rho \partial p_\sigma} &= \frac{8p_\rho p_\sigma - 2(p^2+m^2_A)\delta_{\rho\sigma}}{(m^2_A+p^2)^3}, &
\frac{\partial^2 \Delta_{G}}{\partial p_\rho \partial p_\sigma} &= \frac{8p_\rho p_\sigma - 2(p^2+m^2_G)\delta_{\rho\sigma}}{(m^2_G+p^2)^3}
\label{eq:propG2Derivatives}
\end{align}
for our particular gauge choice. Both integrands are naively divergent, going as  $p^{-4}$, but due to the property $\int \d^4p\, (4p_\mu p_\nu-\delta_{\mu\nu}p^2)f(p^2)=0$, following from $O(4)$ symmetry, they are actually finite.
The only remaining term that can contribute to the wave-function renormalization comes from $\widetilde{\mathcal{M}}^{-1}_{2(A_\mu,G,p)}$, with $m=2$ { in the expansion~\eqref{eq:logExpansion}, {and} $n=0$ in the expansion~\eqref{eq:V21}}, leading to
\begin{equation}
\Gamma \supset -2g^2 \int{\rm d}^4x \frac{{\rm d}^4 p}{(2\pi)^4} \Delta_G\Delta_{A,\mu\nu}(\partial_\mu\varphi)\partial_\nu\varphi.
\end{equation}
All other contributions in Eq.~\eqref{eq:logExpansion} involving higher or mixed powers of $\widetilde{\mathcal{M}}^{-1}_{1/2}$  do not contribute to the wave-function renormalization, as they give rise to interactions with more than two background fields.
With the definitions above we have
\begin{align}\label{eq:Deltaswave}
\int \frac{{\rm d}^4 p}{(2\pi)^4} \Delta_G \Delta_{A,\mu\nu} &= \int \frac{{\rm d}^4 p}{(2\pi)^4} \frac{\delta_{\mu\nu}}{(p^2+m_A^2)(p^2+m_G^2)}\notag\\
&= \int_0^1 {\rm d}w\int\frac{{\rm d}^4 p}{(2\pi)^4} \frac{\delta_{\mu\nu}}{(w(p^2+m_A^2)+(1-w)(p^2+m_G^2))^2}\notag\\
&=\int_0^1 {\rm d}w \int\frac{{\rm d}^3 {\bf p}}{(2\pi)^3} \frac{\delta_{\mu\nu}}{4({\bf p}^2 + w(m_A^2-m_G^2) + m_G^2)^{3/2}}\notag\\
&=\int_0^\infty {\rm d}s \int_0^1 {\rm d}w \int\frac{{\rm d}^3 {\bf p}}{(2\pi)^3} \frac{3\delta_{\mu\nu}}{8({\bf p}^2 + s + w(m_A^2-m_G^2) + m_G^2)^{5/2}}\notag\\
&= \delta_{\mu\nu}\int_0^\infty {\rm d}s \int\frac{{\rm d}^3 {\bf p}}{(2\pi)^3} \frac{1}{4(m_A^2-m_G^2)}\left(\frac{1}{(m_G^2+{\bf p}^2 + s)^{3/2}} - \frac{1}{(m_A^2+{\bf p}^2+s)^{3/2}}\right).
\end{align}
The last manipulations are aimed at defining a kernel containing the divergent part of the effective action as in Eq.~\eqref{eq:Ksdef}. Indeed, defining
\begin{align}\label{eq:Ks}
 K_s({\bf k})\equiv K_{s,(A_4,G)}({\bf k})=\frac{g^2}{(m_A^2-m_G^2)}\left(\frac{1}{(m_G^2+{\bf k}^2 + s)^{3/2}} - \frac{1}{(m_A^2+{\bf k}^2+s)^{3/2}}\right),
\end{align}
we can write
\begin{align}\label{eq:Ks2}\begin{aligned}
\Gamma[\varphi_0] \supset -\frac{1}{2}\int{\rm d}^4x \int_0^{\infty}\D s\frac{{\rm d}^3 p}{(2\pi)^3}K_{s}({\bf k})(\partial_z\varphi_0)^2= -\frac{1}{2}\int{\rm d}^4x\left( \frac{g^2}{4\pi^2}\log(\Lambda^2)(\partial_z\varphi_0)^2+{\rm finite}\right).
\end{aligned}\end{align}
From the divergent contribution we can directly extract the value of the wave-function renormalization constant in the counterterm Lagrangian,
\begin{equation}\label{eq:deltaZwave}
\delta Z = \delta Z_{(A_4,G)}= \frac{g^2}{4\pi^2}\log\frac{\Lambda^2}{\mu^2}.
\end{equation}
By comparison of Eqs.~\eqref{eq:Ks2}, \eqref{eq:deltaZwave} with Eq.~\eqref{eq:Ksdef} we see that indeed $ K_s({\bf k})$ satisfies the desired properties, and it can be used to obtain the renormalized one-loop contributions $B^{(1)}_X$ by means of cutoff-independent integrals, as in Eqs.~\eqref{Gamma:ren}, \eqref{eq:B1ren} and \eqref{eq:B1ren2}. We have identified $\delta Z=\delta Z_{(A_4,G)}, K_s(\bf k)=K_{s,(A_4,G)}(\bf k)$ because, as discussed above, the sector $(A_4,G)$ is the only one that gives rise to divergent wave-function contributions.
\\

\section{Numerical implementation and results}
\label{sec:num}

Given the large amount of time required for the numerical computations, we present results for one set of parameters that illustrate the methods developed in this work.

Before we present the numerical results, we must give some comments.  First, in the thin-wall and planar-wall limits the vacua are assumed to be nearly degenerate and the bubble radius is taken to infinity, with the bounce interpolating between the two vacua (the true vacuum at $z\rightarrow -\infty$, the false vacuum at $z\rightarrow \infty$). Near  $z\rightarrow\pm\infty$, one has quantum fluctuations about a homogeneous background, whose contributions to the effective action are captured by the spacetime integral of the renormalized effective potential. With the effective potential normalized to be zero at the false vacuum, then unless it is exactly degenerate with the true vacuum one expects an infinite contribution coming from the $z$ integral of the renormalized effective potential near $z\rightarrow-\infty$, where the background stays very close to the true vacuum.  This intuition is confirmed quantitatively by our results of the renormalized one-loop contributions to the effective action,  Eqs.~\eqref{eq:B1ren} and \eqref{eq:B1ren2}, which involve the spacetime integral over the one-loop corrections to the effective potential. Adding to this the tree-level action of the bounce, one gets a contribution to the effective action involving the integral over all $z$ of the one-loop renormalized effective potential.
Exact degeneracy between the one-loop energies of the vacua---not just at tree-level---is a necessary requirement for obtaining sensible answers in the planar limit. If the degeneracy at tree-level necessarily implies a $Z_2$ symmetry that exchanges the true and false vacuum, the one-loop degeneracy is automatically satisfied, as it happens for the quartic potentials that were studied previously in the literature \cite{Garbrecht:2015oea,Ai:2018guc}. In those works the tree level degeneracy has been achieved for a vanishing cubic interaction, for which the models exhibit a $Z_2$ symmetry under which the false and true vacua are exchanged.\footnote{Note that in Ref.~\cite{Ai:2018guc}, in which there are Yukawa interactions, the $Z_2$ symmetry involves chiral transformations of the fermion fields.} Our model however lacks this property because, even though one can define a $Z_2$ symmetry, it would relate physically equivalent vacua, rather than the false vacuum at the origin and the true vacuum with spontaneous symmetry breaking. Hence we choose values of the parameters of the tree-level potential $\alpha,\lambda$ and $\lambda_6$ such that for the renormalized Coleman-Weinberg potential of Eq.~\eqref{U:ren:1}, the false and the true vacuum are degenerate. This strict requirement is only due to the planar-wall approximation. Without the latter, the bubble-wall volume is finite and no divergences in the integral of the effective potential inside of the bubble wall appear. We leave the study beyond the planar-wall approximation for future work.

For the numerical evaluation we take $\hbar=1$, for which the effective action is dimensionless, and furthermore we assume appropriate rescaling for the fields and spacetime coordinates that give dimensionless and dimensionful couplings and masses of order one. Note that these rescaling do not affect the value of the effective action. A set of values satisfying the one-loop degeneracy condition is the following:
\begin{equation}
\alpha=2,\quad \lambda_6=\frac{1}{2},\quad\lambda=-2.0254571, \quad g=\frac{1}{2},\quad \mu=\frac{1}{2}.
\label{eq:benchmarkParams2}
\end{equation} 
Note that $\lambda$ has been tuned against the remaining couplings to achieve the degeneracy of the vacua. In the following, to facilitate generalizations beyond the chosen arbitrary units, we will present results for dimensionful quantities in units of the dimensionful parameter $\alpha$. 

In Figure \ref{fig:tuningCW} we show the real part of the one-loop renormalized effective potential in the infinite cutoff limit, obtained from Eq.~\eqref{U:ren:1} and the identities that follow. (For a cutoff  $\Lambda=49=34.65\sqrt{\alpha}$, as will be used below, the total renormalized potential in the region between the vacua differs from its infinite cutoff limit by less than $10^{-13}\alpha^2$). Noting that the symmetry-breaking vacuum appears at $\varphi_-\sim\sqrt{\alpha}=\sqrt{2}$, then given the fact that the tunneling calculations only involve field values $\varphi\leq\varphi_-$, our effective theory treatment with the $|\Phi|^6$ operator will be justified as long as there are UV completions in which higher-dimensional interactions $|\Phi|^{2m}$ with $m>3$---ignored in our calculations---become subdominant for  $\varphi\leq\varphi_-$. To argue that this is the case we can consider a UV completion with heavy Dirac fermions $\Psi,\chi$, in which $\Psi$ is a gauge singlet and $\chi$ has charge -1, so that one can write down a Yukawa coupling
\begin{align}
 {\cal L}_{\rm heavy} \supset -y \bar\Psi\Phi\chi+\rm{c.c.}
\end{align}
For heavy fermion masses of order $M$, then one-loop diagrams induce interactions $\lambda_{2m}|\Phi|^{2m}$ with 
\begin{align}\label{eq:lambdays}
 \lambda_{2m}\sim \frac{y^{2m}}{16\pi^2 M^{2(m-2)}}.
\end{align}
One can quantify the relative impact of the higher-dimensional interactions with $m>3$ for $\varphi\leq\varphi_-\sim\sqrt{\alpha}$ by considering the ratio
\begin{align}\label{eq:ratiointer}
 \left.\frac{\lambda_n|\Phi|^{2m}}{\lambda_6|\Phi|^{6}}\right|_{|\Phi|^2=\alpha}=\left(\frac{y\sqrt{\alpha}}{M}\right)^{2(m-3)}.
\end{align}
In our benchmark scenario, we have that Eq.~\eqref{eq:benchmarkParams2} and \eqref{eq:lambdays} imply $M\sim{y^3}/{2 \sqrt{2} \pi }$ such as to generate a value of $\lambda_6$ of the assumed size. Substituting this into Eq.~\eqref{eq:ratiointer}, and imposing $\alpha=2$ we get
\begin{align}\label{eq:ratiointer2}
 \left.\frac{\lambda_n|\Phi|^{2m}}{\lambda_6|\Phi|^{6}}\right|_{|\Phi|^2=\alpha=2}=\left(\frac{4\pi}{y^2}\right)^{2(m-3)}.
\end{align}
Thus we can get a relative suppression of $(1/10)^n$ for every $|\Phi|^{6+2n}$ interaction for $y=2 \sqrt[4]{10} \sqrt{\pi }$, still within the perturbative bound $y<4\pi$. Thus, for this example, it is consistent to ignore the higher-dimensional operators beyond $|\Phi|^6$ in our analysis.

As we are forced to consider the degenerate limit of the one-loop potential, the simplest way to proceed with the numerical determination of the Green's functions is to use as initial background $\varphi_0$  the  solution to the equation of motion using the real part of the {one-loop} effective potential. In the thin-wall approximation appropriate for nearly degenerate vacua, implying a large bubble radius, the bounce is computed by neglecting the friction term appearing in Eq.~\eqref{eq:eomExpanded}. Substituting the tree-level potential by its one-loop counterpart, and implementing the planar wall approximation by substituting $r$ with  $z$, we have to solve
\begin{align}
-\frac{\d^2\varphi}{\d z^2} +{\rm Re} (U^{\rm ren}_{\rm CW})'(\varphi) =&\, 0, \quad \left.\varphi_0(z)\right|_{z\rightarrow\infty}=0,\,\, \left.\varphi_0'(z)\right|_{z\rightarrow -\infty}=0.
\label{eq:eomExpanded2}
\end{align}
The solution is found numerically and is adjusted so that the wall location, defined as the point where the derivative is maximal, is located at $z=0$. In terms of the compact variable $u$ of Eq.~\eqref{eq:uz}---which will be used in the remainder of this section---the initial bounce $\varphi_0$ is shown as the dashed orange line in Figure~\ref{fig:bounceCorrections}.

This bounce is then used as the background for the remainder of the numerical analysis. {We note that since there is translation symmetry along the directions parallel to the bubble wall, all the quantities $B$, $B^{(1)}$ etc. are proportional to $V=\int{\rm d}^3{\bf x}$ (cf. Eq.~\eqref{eq:functionalDetDeformation}). In the following, all the quantities are understood with this three-volume factored out. 
\begin{figure}[htbp]
\centering
\includegraphics[width=0.7\textwidth]{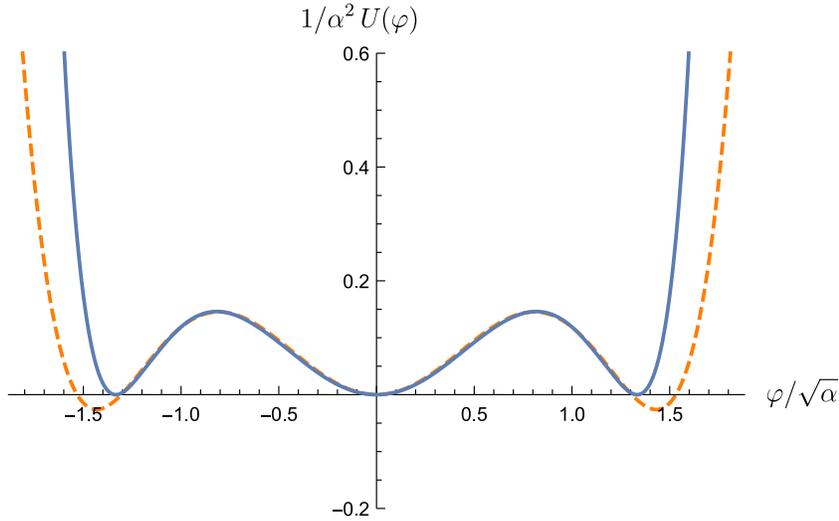}
\captionsetup{justification=raggedright,singlelinecheck=false}
\caption{Tree-level potential (orange dashed) and renormalized Coleman--Weinberg potential (solid blue) with enforced degeneracy between the vacua. The values used for the couplings are stated in Eq.~\eqref{eq:benchmarkParams2}.}
\label{fig:tuningCW}
\end{figure}

\begin{figure}
	\begin{center}
	\includegraphics[width=.49\textwidth]{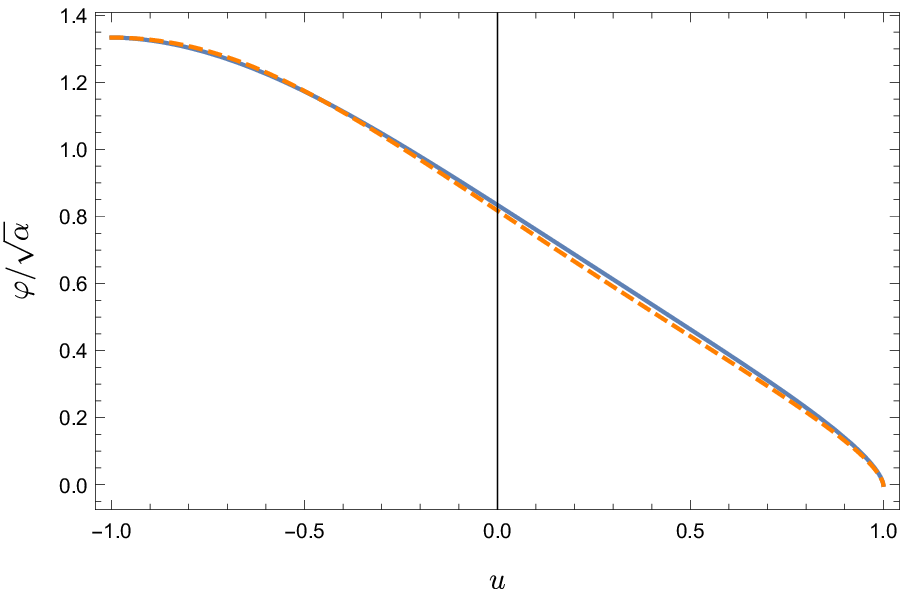}%
	\hfill\includegraphics[width=.485\textwidth]{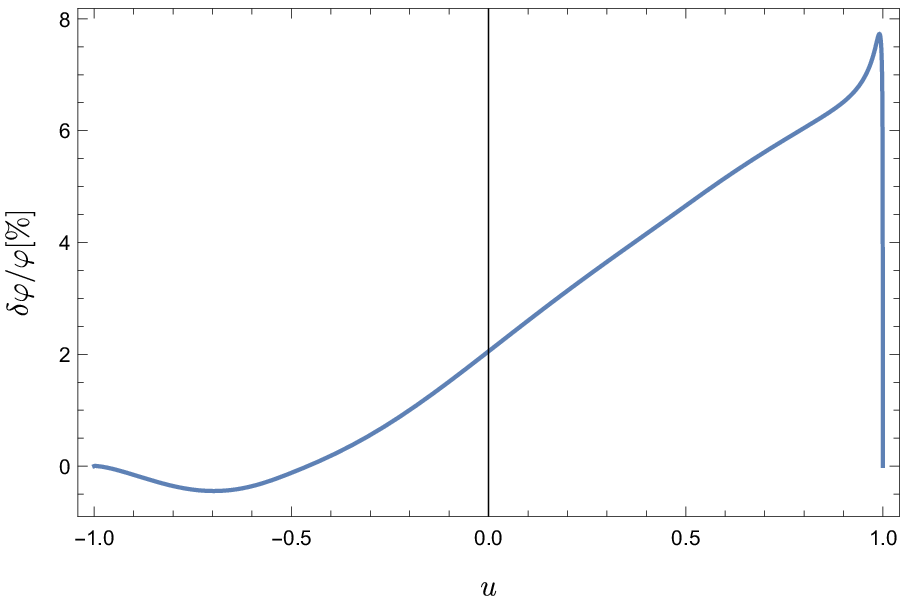}
	\end{center}
	\captionsetup{justification=raggedright,singlelinecheck=false}
	\caption{Left: Initial approximation to the bounce (dashed orange) and the  version including gradient corrections arising from the self-energies computed above (solid blue). Right: relative variation of the bounce induced by gradient corrections.}
	\label{fig:bounceCorrections}
\end{figure}

The code which solves for the Green's function of the gauge-Goldstone block is run for a range of values of the three-momentum {$|{\bf k}|$} ranging from $0$ to $|{\bf k}_{\rm max}|=50$. For $|{\bf k}|<0.5$ we use the direct method of Section~\ref{subsec:exactmethod}, while for larger $|{\bf k}|$ we use the iterative approach described in \ref{subsec:perturbativemethod}. For the latter, we have adapted the number of iterations depending on the value of {$|{\bf k}|$}: lower values of {$|{\bf k}|$} need more iterations to converge, while for higher values the solution stabilizes faster. Once the iterations are completed and the solutions for the range of $|\mathbf{k}|$ compiled and once the solutions in the $\hat\Phi$ and $(\bar\eta,\eta)$ sectors are also calculated, the coincident limit is taken in order to compute quantities such as the resolvents ${\cal M}_{X;\sqrt{{\bf k}^2+s}}(\varphi_0)$ and the renormalized determinants $B^{(1){\rm ren}}_X$, see Eqs.~\eqref{eq:B1ren}, \eqref{eq:B1ren2}.} The numerically generated solutions allow us to integrate up to values of $|{\bf k}|=\Lambda\leq 49=34.65\sqrt{\alpha}$ for the momentum along the wall directions. The $s$ integration in \eqref{eq:B1ren2} has been constructed such that it remains finite for a large cutoff $\Lambda$ in the $|\bf k|$ integration, so that the dependence on the upper limit of the $s$ integration is suppressed; we have checked this explicitly by comparing results in which we integrate up to a value of $s=\Lambda_s^2>\Lambda^2$, and in which we extrapolate the integrand for large $s$ (after performing the integrals in $u,|\bf k|$) using a power-law fit, and integrate up to $s=\infty$. The results agree within percent precision or better. From the $B^{(1){\rm ren}}_X$ one can obtain the contribution ${\cal B}^{(1)\rm ren}$ to the renormalized effective action using Eq.~\eqref{Gamma:ren}.

The renormalized tadpoles $(\Pi_X(\varphi_0)\varphi_0)^{\rm ren}$ follow then from Eqs.~\eqref{eq:Piphiall}, \eqref{eq:tapoleA4Gauge}, \eqref{eq:Piren},  and from them one readily obtains  the correction to the bounce $\delta\varphi^{\rm ren}$ and  the renormalized two-loop correction to the effective action ${\cal B}^{(2){\rm ren}}$ (see  Eqs.~\eqref{eq:deltaB2ren}, \eqref{eq:B2rentotal}). 
  
A subtlety in the calculation of the $|\bf k|$ integrations is the appearance of an integrable singularity in the $X=\Phi$ sector, which arises from a divergence of the resolvent ${\cal M}_{\hat\Phi;\sqrt{{\bf k}^2+s}}$ of the form
\begin{align}\label{eq:Mdiv}
{\cal M}_{\hat\Phi;\sqrt{{\bf k}^2+s}}(\varphi_0;z,z)\sim \frac{\varphi_-(z)^2}{{\bf k}^2+s-\lambda_-}+O(({\bf k}^2+s-\lambda_-)^0)\text{     for      }\lambda_-=0.21933.
\end{align}
The reason for such a divergence is simply that the operator ${\cal M}^{-1}_{\hat\Phi;{\bf k}=0}(\varphi_0)$, when acting on functions of $z$, has a discrete negative mode $\varphi_-(z)$ with eigenvalue $-\lambda_-$, i.e.
\begin{align}\label{eq:negeigen}
{\cal M}^{-1}_{\hat\Phi;{\bf k}=0}(\varphi_0)\varphi_-(z) = -\lambda_-\varphi_-(z).
\end{align}
The existence of such a negative mode can be understood from the fact that, although we have enforced degeneracy of the potential at the one-loop level, the tree-level vacua are not degenerate (see Fig.~\ref{fig:tuningCW}). As is well known from one-dimensional tunneling calculations, the fluctuation operator in the background of a configuration that interpolates between  non-degenerate vacua has a negative mode, and from the existence of the latter and the spectral decomposition of the resolvent one infers a contribution of the form of Eq.~\eqref{eq:Mdiv}. Luckily, such contribution still leads to convergent integrals in $|\bf{k}|$ in the evaluation of the determinant contributions $g_X B^{(1)}_X$ and the tadpoles $g_X\Pi_X(\varphi_0)\varphi_0$, because the integral of ${\bf k}^2/({\bf k}^2-b^2)$ around $|{\bf k}|=b$ is finite:
\begin{align}\label{eq:intkb}
 \int_{b-\delta}^{b+\delta}\d|{\bf k}| \frac{{\bf k}^2}{{\bf k}^2-b^2}=\frac{1}{2} \left(4 \delta +b \log \left(\frac{2 b-\delta}{2b+\delta}\right)\right).
\end{align}
To avoid numerical instabilities we treat separately the integration of the resolvent (or Green's function, in the case of tadpoles) around the singularity, expressing it as a contribution coming from the difference between the resolvent and the divergent piece of Eq.~\eqref{eq:Mdiv}--with $\varphi_-(z)$ and $\lambda_-$ computed numerically by solving the eigenvalue equation \eqref{eq:negeigen}---plus the contribution from the divergent piece alone. The first term yields a finite result, while the $|\bf k|$ integral of the divergent term is calculated using the analytic result of Eq.~\eqref{eq:intkb}, i.e. it is evaluated in the principal value sense.

As follows from the arguments at the beginning of this section, in the considered limit of exact degeneracy between the one-loop energies of the vacua (which in our case implies nondegeneracy at tree level), only the combination $({\cal B}^{(0)}+{\cal B}^{(1)\rm ren})/V$ is finite, and thus we will not fully distinguish between the  two contributing terms.  Rather, we separate homogeneous and gradient contributions using  Eqs.~\eqref{eq:B1ren}, \eqref{eq:B1ren2} and~\eqref{Gamma:ren}, and write
\begin{align}\begin{aligned}
{\cal B}^{(1)\rm ren}=&\,{\cal B}^{(1)\rm ren, hom} +{\cal B}^{(1)\rm ren,grad},
\end{aligned}\end{align}
where
\begin{align}\begin{aligned}
{\cal B}^{(1)\rm ren,hom} =&\,\sum_X g_X B^{(1)\rm ren, hom}, & {\cal B}^{(1)\rm ren,grad} =&\,\sum_X g_X B^{(1)\rm ren, grad}.
\end{aligned}\end{align}
It follows that
\begin{align}
\frac{1}{V}({\cal B}^{(0)}+{\cal B}^{(1)\rm ren,hom})=\int dz\left(\frac{1}{2}(\partial_z\varphi_0(z))^2+{\rm Re}\,U^{\rm ren}_{\rm CW}(\varphi_0)\right),
\end{align}
which involves the integral over the full one-loop potential. As a consequence of Eq.~\eqref{eq:eomExpanded2}, our initial bounce $\varphi_0$ is chosen to extremize the above combination $({\cal B}^{(0)}+{\cal B}^{(1)\rm ren,hom})/V$, and the result is finite. The renormalized gradient contribution ${\cal B}^{(1)\rm ren,grad}$ likewise involves no pieces that are divergent under spacetime integration and remains finite.

\begin{figure}
\centering
\includegraphics[width=0.495\textwidth]{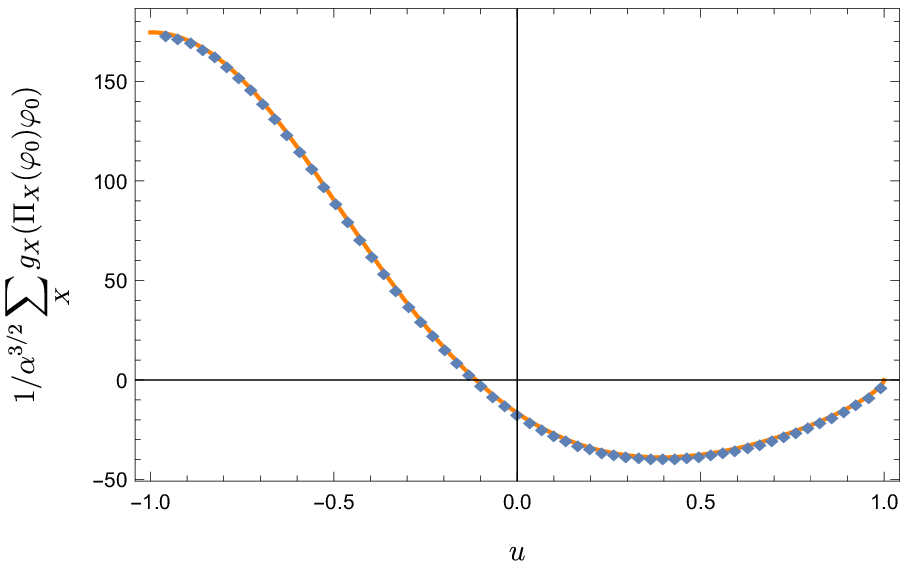}\hfill%
\includegraphics[width=0.495\textwidth]{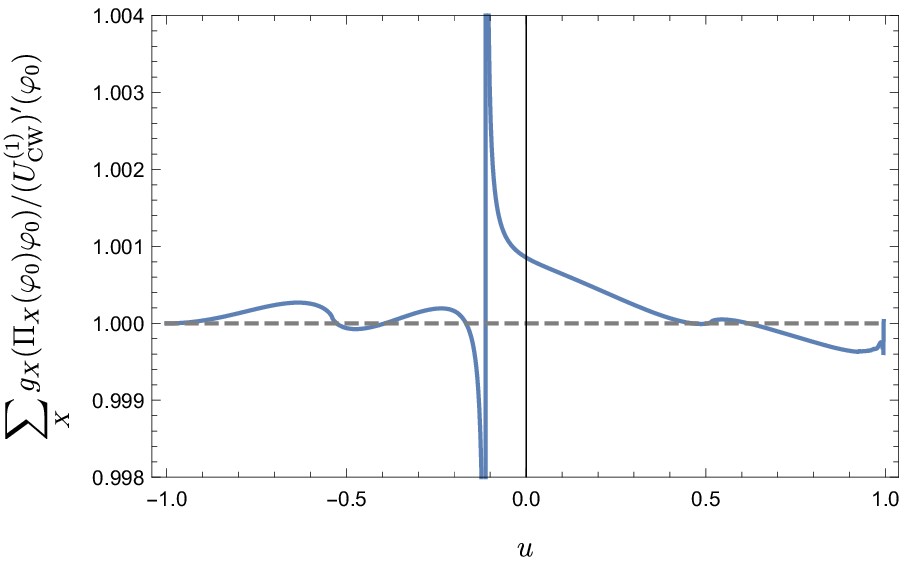}
\captionsetup{justification=raggedright,singlelinecheck=false}
\caption{Left: tadpole $\sum_X\Pi_X(\varphi_0(u);u)\varphi_0(u)$ with gradient effects  (diamonds) and tadpole $(U^{(1)}_{\rm CW})'(\varphi_0)=$ $\sum_X\Pi_{X;{\rm hom}}(\varphi_0(u);u)\varphi_0(u)$  without gradient effects (solid). Right: Ratio of the total tadpole contribution over its counterpart without gradient effects. The cutoff is taken as $\Lambda=34.65\sqrt{\alpha}$. }
\label{fig:tadpoleVainilla}
\end{figure}

\begin{figure}[htbp]
	\centering
	\includegraphics[width=.493\textwidth]{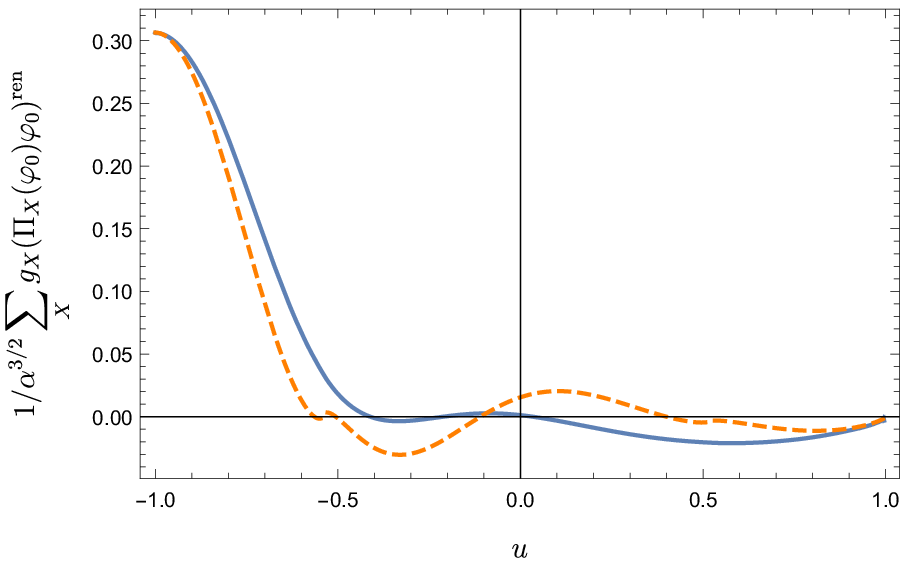}\hfill\includegraphics[width=.497\textwidth]{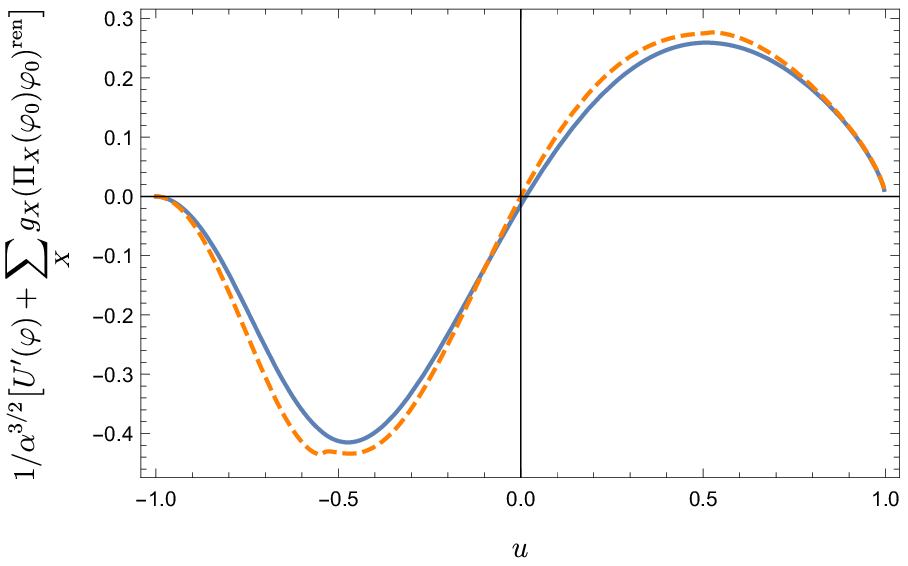}
	\captionsetup{justification=raggedright,singlelinecheck=false}
	\caption{Left: Plot of the total renormalized tadpole $\sum_X(\Pi_X(\varphi_0(u);u)\varphi_0(u))^{\rm ren}$  (solid blue line), and its approximation neglecting gradients, $(U^{(1){\rm ren}}_{\rm CW})'(\varphi_0)=$ $\sum_X(\Pi_{X;{\rm hom}}(\varphi_0(u);u)\varphi_0(u))^{\rm ren}$ (dashed orange), as a function of the compactified radial coordinate $u$.  Right: Analogous plot, adding the tree-level tadpole contribution $U'(\phi)$ so as to obtain the one-loop functional derivative of the effective action at the bounce $\varphi_0$. The cutoff is taken as $\Lambda=34.65\sqrt{\alpha}$.}
	\label{fig:tadpole+CW}
\end{figure}

\begin{figure}[htbp]
	\centering
	\includegraphics[width=.495\textwidth]{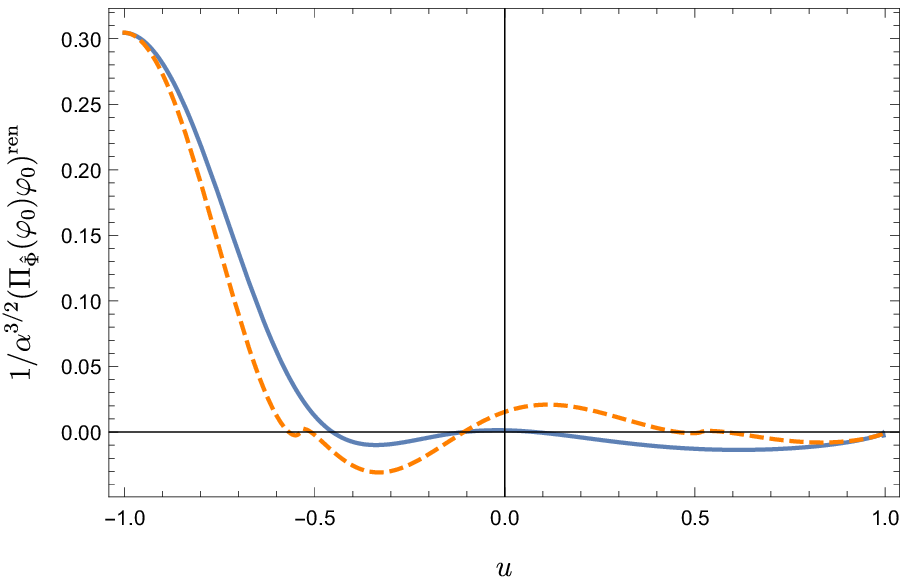}
	\hfill\includegraphics[width=.495\textwidth]{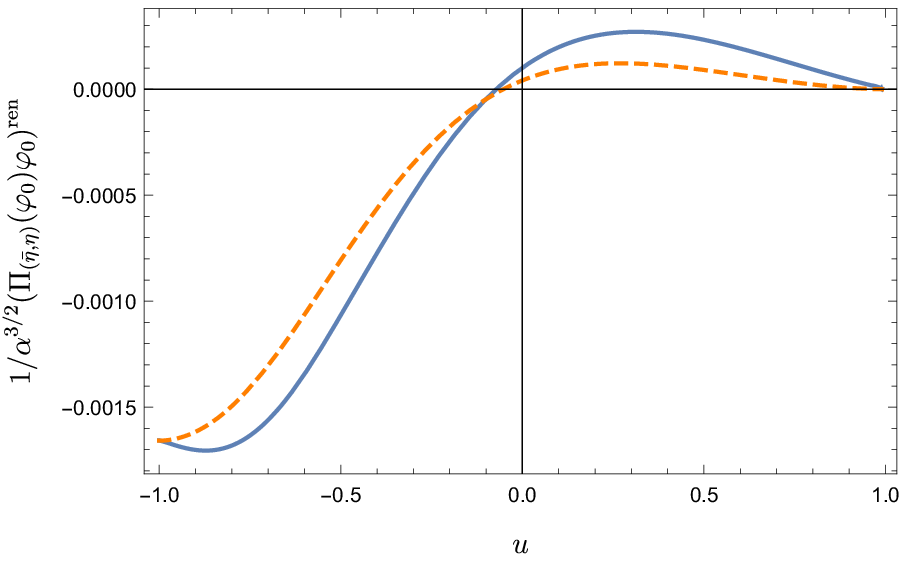}
	\includegraphics[width=.495\textwidth]{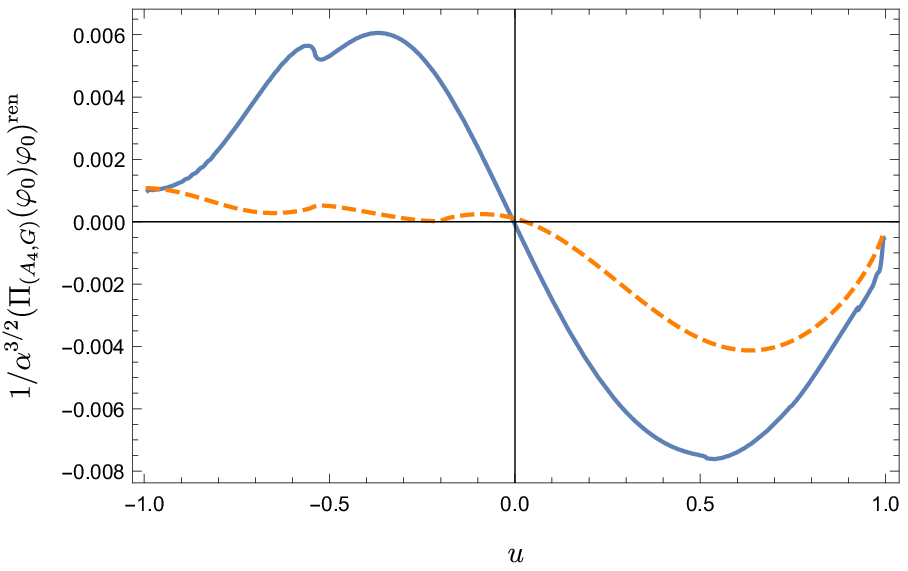}\hfill\includegraphics[width=.495\textwidth]{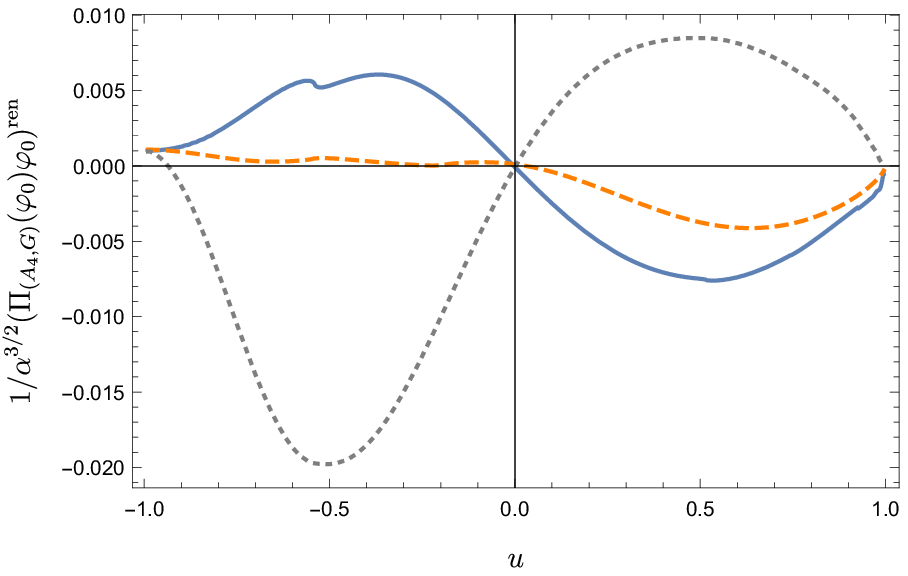}
	\captionsetup{justification=raggedright,singlelinecheck=false}
	\caption{For  $\Lambda=34.65\sqrt{\alpha}$, renormalized tadpoles $(\Pi_X(\varphi_0(u);u)\varphi_0(u))^{\rm ren}$ for each set of fields (solid blue) and the corresponding terms ignoring gradient effects, $(U^{(1)}_{\rm CW,X})'(\varphi_0)=$ $(\Pi_{X;{\rm hom}}(\varphi_0)\varphi_0)^{\rm ren}$ (dashed orange). From upper left to lower right we have $X=\hat\Phi,(\bar\eta,\eta),(A_4,G), (A_4,G)$. The lower right plot includes an extra dotted grey curve illustrating the result in the $(A_4,G)$ sector when one ignores wave-function renormalization. Note that this latter curve is cutoff dependent because the logarithmically divergent wave-function renormalization has not been subtracted.}
	\label{fig:tadpoleRenPerField}
\end{figure}

Figure~\ref{fig:tadpoleVainilla} shows the numerical result for the total tadpole, $\sum_X g_X\Pi_X(\varphi_0;u)\varphi_0(u)$,  before renormalization, obtained using equations~\eqref{eq:PiPhi}, \eqref{eq:PiEta} and~\eqref{eq:tapoleA4Gauge},  compared with the corresponding result using the homogeneous Green's functions \eqref{eq:Pihom}, which do not include gradient effects. The similarity between both calculations is due to the fact that, prior to renormalization, the leading cutoff-dependent contributions dominate, and they are fully captured by the homogeneous Green's functions. In contrast to this, Figure~\ref{fig:tadpole+CW} shows the sum of the renormalized tadpoles with  (solid blue) and without (dashed orange) gradient effects, whose impact becomes now manifest after the subtraction of the cutoff dependence. For comparison, we also show
in Figure~\ref{fig:tadpoleRenPerField} the renormalized tadpoles for the individual
sectors $X$, whose $g_X$-weighted sum constitutes the  quantity shown in Figure~\ref{fig:tadpole+CW}. From Figures~\ref{fig:tadpoleRenPerField} and \ref{fig:tadpole+CW} it is clear that scalar fluctuations dominate the total result, and that the sectors more affected by gradient corrections are those involving the degrees of freedom $\hat\Phi$ and $(A_4,G)$. The scalar dominance, seen both for the homogeneous and full results, is due to the large ratio between the scalar quartic and the gauge coupling in our benchmark scenario (see Eq.~\eqref{eq:benchmarkParams2}).
The  gradient corrections are typically of order 100\% of the homogeneous results at the same loop order, and even larger for $X=(A_4,G)$. As the tadpoles are one-loop quantities, it follows that one-loop gradient effects can become equally (or more) important than homogeneous one-loop effects. Note that in the $(A_4,G)$ sector it is important to include the effect of wave-function renormalization, as if it were ignored one would obtain much larger gradient corrections, as shown in the dotted grey line in the lower right plot of Figure~\ref{fig:tadpoleRenPerField}.  In the right plot of Figure~\ref{fig:tadpole+CW} we show the quantity $U'(\varphi_0)+\sum_X(\Pi_X(\varphi_0)\varphi_0)^{\rm ren}$, which corresponds to the functional derivative of the full one-loop effective action evaluated at the bounce. Note that it approaches zero at $u=\pm1$, i.e. when the field reaches the two vacua. As the latter are extrema of the one-loop effective action, one indeed expects $\delta\Gamma/\delta\varphi=0$. In general, as for $u=\pm1$ ($z=\pm\infty$) the field approaches the vacua with zero derivative with respect to $z$, one expects gradient effects to go to zero, which is indeed observed in Figures~\ref{fig:tadpoleRenPerField} and \ref{fig:tadpole+CW}. 
These tadpole contributions lead to the quantum-corrected bounce shown as a solid blue line in the left plot in Figure~\ref{fig:bounceCorrections}, to be compared with the initial approximation shown in dashed orange. As the initial bounce solves the equations of motion for the one-loop effective action in the homogeneous approximation, the correction to the bounce shown on the right plot of Figure~\ref{fig:bounceCorrections} is purely due to gradient effects, and can be seen to stay below a few percent. These corrections are somewhat larger than the effects  found in the studies of Refs.~\cite{Garbrecht:2015oea,Garbrecht:2015yza,Ai:2018guc}. In those works, the models have an emerging $Z_2$ symmetry in the limit of degenerate vacua which implies a negative parity symmetry for the bounce and its corrections, (i.e. $\varphi(-u)=-\varphi(u)$ so that the tadpoles vanish at $u=0$ because $\varphi(0)=0$), that ultimately constrains gradient corrections to be zero around $u=0$, where one would naively expect maximal effects due to the larger derivative of the bounce. Such a symmetry is not present here, which may explain  the larger effects. Again, gradient corrections go to zero for $u=\pm1$. As the bounce accounts for both tree and loop-level effects, and a relative loop factor with order one couplings is expected to lead to percent corrections, it follows that gradient corrections to the bounce are roughly of the same order (or greater) as generic one-loop effects, which matches what was seen in the renormalized tadpoles.

With the above results for the renormalized tadpoles and for the correction of the bounce configuration, we can finally estimate the contributions ${\cal B}^{(i)}$ for the effective action appearing in the decay rate~\eqref{eq:expandedDecayRate}. These are given in  Table~\ref{tab:determinantTotals}. Table~\ref{tab:determinantResults} gives the one-loop contributions $g_X B^{(1)\rm ren,grad}_X$ arising from gradient effects, together with their percentual weight on the total one-loop contribution ${\cal B}^{(0)}+{\cal B}^{(1)\rm ren}$. Again, gradient corrections are of order 1\% of the tree-level plus one-loop result, i.e. the same size of generic one-loop effects. All the individual contributions have a positive sign, as expected in general for bosonic loop corrections.
\begin{table*}[htbp]
	\makegapedcells
	\begin{tabularx}{.45\textwidth}{c|c}
		\hline
		&\quad Value [$\times\alpha^{-3/2}$] \\
		\hline
		\hline	
		$({\cal B}^{(0)}+{\cal B}^{(1)\mathrm{ren}})/V$ & $0.473$ \\
		\hline
		${\cal B}^{(2)\mathrm{ren}}/V$ & $-0.000345$ \\
		\hline
		$({\cal B}^{(0)} + {\cal B}^{(1)\mathrm{ren}} + {\cal B}^{(2)\mathrm{ren}})/V$ & $0.474$\\
		\hline
	\end{tabularx}
	\caption{Numerical results for the renormalized contributions to the effective action.}
	\label{tab:determinantTotals}	
\end{table*}

\begin{table*}[htbp]
	\makegapedcells
	\begin{tabularx}{.77\textwidth}{c|c|c}
	\hline
	 &\quad Value [$\times\alpha^{-3/2}$]\hspace*{3mm} & \quad  ${\rm Value}/(({\cal B}^{(0)}+{\cal B}^{(1)\rm ren})/V)$ \hspace{1mm} [$\%$]\\
	\hline
	\hline	
	$ g_{\hat\Phi}B^{(1)\rm ren,grad}_{\hat{\Phi}}/V$ & 0.00139 & 0.29 \\
	\hline
	$ g_{(\bar\eta,\eta)}B^{(1)\rm ren,grad}_{(\bar\eta,\eta)}/V$ & 0.0000748 & 0.016 \\
	\hline
	$g_{(A_4,G)}B^{(1)\rm ren,grad}_{(A_4,G)}/V$ &0.00332 & 0.70\\
	\hline
	$\sum_X g_{X}B^{(1)\rm ren,grad}_{X}/V$ & 0.00479 & 1.0\\
	\hline
	\end{tabularx}
	\caption{Numerical results for the gradient contributions to the determinant terms of the effective action.} 
\label{tab:determinantResults}	
\end{table*}

It should be noted that the relative weight of the one-loop and two-loop corrections is in accordance with the expectations of perturbation theory with order one couplings. Aside from the already mentioned fact that the one-loop gradient contributions in Table \ref{tab:determinantResults} are two orders of magnitude (roughly a loop factor) below the tree-level plus one-loop result ${\cal B}^{(0)}+{\cal B}^{(1)\rm ren}$ in Table \ref{tab:determinantTotals}, one also has that the former is four orders of magnitude above the two-loop contribution ${\cal B}^{(2)\rm ren}$.

\section{Conclusions}
\label{sec:Conc}
The self-consistent Green's function method of Ref.~\cite{Garbrecht:2015oea} for the calculation of radiative corrections to decay rates of false vacuum states, which allows to account for all gradient effects at a given loop order, has been applied here for the first time to a gauge theory.  We have considered a $U(1)$ gauge field coupled to a complex scalar and, in order to have two physically distinct vacua amenable to tunneling transitions at zero temperature, we have considered a potential including a higher-dimensional $|\Phi|^6$ interaction. The model is intended as an illustration of how the method of  Ref.~\cite{Garbrecht:2015oea} applies to gauge theories, as a first step on the way towards self-consistent calculations of vacuum decay in theories like the SM. The model studied in this article  can be considered as an effective description of a UV theory in which heavy fermions have been integrated out. Our specific choice of parameters is consistent with a UV completion in which higher-order interactions $|\Phi|^{2m}, m>3$, are subdominant, justifying our truncation beyond $|\Phi|^6$.

In the limit of degenerate vacua, leading to a planar thin-wall regime, we have included corrections to the effective action coming from background gradients and from the shift of the background induced by quantum effects. As expected from the bosonic nature of the gauge and scalar fields, quantum corrections to the effective action are positive,  leading to a longer lifetime for the false vacuum. Our results also show that gradient corrections are of the same order as homogeneous one-loop corrections. This implies that considering only the leading terms in a gradient expansion at one loop would result in theoretical uncertainties that would remain of the order of a loop factor. Hence, accounting for full gradient effects is crucial to achieving full one-loop accuracy. The method applied here captures all one-loop effects plus two-loop corrections associated with dumbbell diagrams. The latter can be the dominant two-loop effects in more general models with more degrees of freedom, e.g. non-Abelian theories or in the presence of several spectator fields \cite{Ai:2018guc}.

In relation to previous applications of the Green's function method to tunneling calculations in the thin-wall regime~\cite{Garbrecht:2015oea,Garbrecht:2015yza,Ai:2018guc}, in which gradient effects were found to be comparable to two-loop corrections, here we have found comparatively larger gradient corrections. This  can be due to the fact that, in contrast to the case of the aforementioned works, in the model studied here there is no emergent $Z_2$ symmetry, which exchanges the false and true vacua, in the limit of degenerate vacua. In earlier works such symmetry led to parity constraints in the $z$-dependence of the gradient corrections, which limited their impact. On the other hand, the use of the Green's function method has already been shown to have an important effect on the results for a  non-$Z_2$ symmetric setup  away from the thin-wall limit, as in the scale-invariant scalar model of Ref.~\cite{Garbrecht:2018rqx}.

In comparison to earlier applications of the Green's function method for tunneling computations, the present work has addressed the following novel challenges:

\begin{itemize}
 \item The gauge and Goldstone boson fluctuations form a coupled system, for which earlier methods to compute the Green's functions are no longer applicable. In contrast to the usual calculations in a constant background in gauge theories, the freedom in the gauge fixing procedure does not allow to eliminate the mixing in the presence of background gradients. Nevertheless, a judicious choice of gauge-fixing allows to restrict the mixing so that, in the planar limit, it only involves a single gauge field component and the Goldstone degree of freedom of the complex scalar. At low values of the momenta in the directions parallel to the wall one can solve directly for the full mixed equations numerically, while for larger momenta one is forced to use an iterative method to account for the mixing effects.
 
 \item The planar thin-wall regime can be reached by enforcing degeneracy between the two vacua. However, as mentioned before and in contrast to previous studies, this limit is not associated with an emergent $Z_2$ symmetry enforcing degeneracy of the vacua at every order of perturbation theory. In particular, this means that starting with a potential with tree-level degeneracy leads to long-distance divergences in the quantum corrections to the bounce action, which arise from the mismatch between the vacuum energies at one-loop. This simply means that planar-wall approximation is no longer applicable at one loop unless one enforces vacuum degeneracy at one loop rather than at tree level and  uses an initial bounce configuration which solves the Euclidean equations of motion corrected with the one-loop Coleman-Weinberg potential. Finiteness of the bounce action at one loop is achieved when considering tree and loop effects jointly.
\end{itemize}

The techniques developed here can be applied in future studies on models that capture more of the features present in the case of the SM, e.g. going beyond the thin-wall approximation, or considering non-Abelian gauge theories. Moreover, it would be of interest to directly evaluate the sensitivity of the quantum corrections to the tunneling rate with respect to changes in the gauge-fixing parameters.\\

{\it Acknowledgments}
This work has been supported in part by SFB 1258 of the Deutsche Forschungsgemeinschaft.

\renewcommand{\addcontentsline}[3]{}
\renewcommand{\section}{}
\renewcommand{\refname}{}

\bibliography{references}

\end{document}